\documentclass[]{spie} 

\usepackage{graphicx}
\usepackage[flushleft]{threeparttable}
\usepackage{amsfonts}
\usepackage{amsmath}
\usepackage{multirow}
\usepackage{makecell}
\usepackage[colorlinks=true, allcolors=blue]{hyperref}
\usepackage{subcaption}
\usepackage{hhline}
\usepackage{booktabs}
\usepackage[ruled,vlined]{algorithm2e}
\usepackage{bm}
\usepackage{setspace}

\graphicspath{ {.} }

\newcommand{\etal}{\textit{et al}. }
\title{Joint localization and classification of breast tumors on ultrasound images using a novel auxiliary attention-based framework}

\author[a]{Zong Fan}
\author[b]{Ping Gong}
\author[b]{Shanshan Tang}
\author[b]{Christine U. Lee}
\author[a]{Xiaohui Zhang}
\author[a,c,e,f]{Pengfei Song}
\author[b]{Shigao Chen}
\author[a,d,e]{Hua Li}
\affil[a]{Department of Bioengineering, University of Illinois at Urbana-Champaign, IL, USA}
\affil[b]{Mayo Clinic, Rochester, Minnesota, USA}
\affil[c]{Department of Elect. \& Computer Eng., University of Illinois at Urbana-Champaign, IL, USA}
\affil[d]{Department of Radiation Oncology, Washington University in St. Louis, MO, USA}
\affil[e]{Cancer Center at Illinois, Urbana, IL, USA}
\affil[f]{Beckman Institute, University of Illinois at Urbana-Champaign, IL, USA}

\authorinfo{Send correspondence to Hua Li. E-mail: li.hua@wustl.edu}

\begin{document}
\maketitle
\begin{abstract}
Automatic breast lesion detection and classification is an important task in computer-aided diagnosis,
in which breast ultrasound (BUS) imaging is a common and frequently used screening tool.
Recently, a number of deep learning-based methods have been proposed for joint localization and classification of breast lesions using BUS images.
In these methods, features extracted by a shared network trunk are appended by two independent network branches to achieve classification and localization.
Improper information sharing might cause conflicts in feature optimization in the two branches and leads to performance degradation.
Also, these methods generally require large amounts of pixel-level annotated data for model training.
To overcome these limitations,
we proposed a novel joint localization and classification model based on the attention mechanism and disentangled semi-supervised learning strategy.
The model used in this study is composed of a classification network and an auxiliary lesion-aware network.
By use of the attention mechanism, the auxiliary lesion-aware network can optimize multi-scale intermediate feature maps and extract rich semantic information to improve classification and localization performance. 
The disentangled semi-supervised learning strategy only requires incomplete training datasets for model training.
The proposed modularized framework allows flexible network replacement
to be generalized for various applications.
Experimental results on two different breast ultrasound image datasets demonstrate the effectiveness of the proposed method.
The impacts of various network factors on model performance are also investigated to gain deep insights into the designed framework.
\end{abstract}


\section{Introduction}
\label{sec:intro}

Breast cancer is the most frequent cause of death in women aged between 35-55 years~\cite{pyakurel2014study,nothacker2009early}.
Ultrasonography screening is a common tool for early diagnosis of breast lesions due to its cost-effectiveness and safety~\cite{lee2010breast}.
During the past years, a number of computer-aided diagnosis methods have been proposed to assist in lesion localization and classification.
These automatic screening methods range from conventional machine learning techniques~\cite{Gu2016Automated3U,shan2016computer,Xu2019MedicalBU}, to deep learning (DL) techniques~\cite{wu2019artificial,yap2018breast,shin2018joint}.
Particularly, DL-based methods have achieved great success due to their powerful learning capabilities~\cite{shin2018joint,Han2017ADL}.
Generally, the localization task is represented as either lesion segmentation or detection.
The segmentation task aims to accurately delineate the lesion regions~\cite{vakanski2020attention,tang2021feature},
while the detection task is to simply predict lesion locations in the form of bounding boxes~\cite{cao2017breast,shin2018joint},
DL-based classification methods are developed to stratify lesions into subgroups to help clinicians design appropriate treatment strategies.  

Multi-task learning (MTL) models have been proposed to conduct these two tasks simultaneously to increase data efficiency without sacrificing the performance of each task.
They generally consist of a shared feature extractor with two appended task-specific branches.
Given the discriminative features extracted by the shared feature extractor, 
the classification branch differentiates lesion types
and the localization branch confines the potential lesion regions, respectively~\cite{Crawshaw2020MultiTaskLW}.
This shared design leverages semantic information to decode the lesion type and location simultaneously, 
which can reduce the risk of overfitting and improve learning efficiency and robustness~\cite{Crawshaw2020MultiTaskLW,zhou2016learning,rasaee2021explainable,Zhou2021MultitaskLF,Chowdary2022AML,Zhang2021SHAMTLSA,Xu2022MultiTaskLW}.
Zhou \etal employed an encoder-decoder network (VNet) for the segmentation task, 
while the intermediate feature maps were reused for classification by a lightweight network only consisting of a global average pooling layer and three fully-connected layers~\cite{Zhou2021MultitaskLF}.  
Chowdary \etal employed residual U-Net architecture for segmentation and shared the intermediate feature maps for classification with a two-layer fully-connected (FC) network~\cite{Chowdary2022AML}.
Some MTL methods simplify pixel-wise lesion segmentation to detection in the form of bounding boxes instead~\cite{cao2017breast,shin2018joint}.
For instance, Cao \etal studied and compared the performance of several popular object detection methods such as YOLO and SSD~\cite{cao2017breast}.
Shin \etal employed Faster-RCNN for detection and classification of breast tumors on a BUS image dataset~\cite{shin2018joint}.

In these traditional MTL methods, balancing the degree of the information shared between the two different tasks is critical to ensure the model performance~\cite{li2022task}. 
Improper information sharing may decrease the model performance due to the conflicts in optimizing extracted features between the two tasks with different objectives~\cite{Crawshaw2020MultiTaskLW}.
Loss weighting is a common method that balances and tunes the individual loss functions for different tasks~\cite{Crawshaw2020MultiTaskLW,Liu2019EndToEndML}.
Liu \etal proposed an adaptive weighting method to dynamically balance the learning rate of each task~\cite{Liu2019EndToEndML}.
Gradient demodulation methods modify training gradients to alleviate the conflicts of learning dynamics between different tasks~\cite{Crawshaw2020MultiTaskLW, sinha2018gradient}. 
Sinha \etal employed adversarial training to align the gradients from different tasks to boost the model performance~\cite{sinha2018gradient}.
Also, the attention mechanism is widely employed to consider the correlation of different tasks to improve model learning capability and performance ~\cite{Xu2022MultiTaskLW,Zhang2021SHAMTLSA,MA2020101800,singh2020breast}. 
This technique enables the extracted features to focus on more discriminative information.
Xu \etal proposed a self-attention module on top of a U-Net to utilize the context information to improve both breast tumor segmentation performance and classification performance~\cite{Xu2022MultiTaskLW}.

Conventional DL-based methods for detection and classification of breast lesion typically require large amounts of fully-annotated training images,
which is very time- and effort-consuming.
Semi-supervised learning techniques can alleviate the burden of annotating localization labels, 
which automatically exploit incomplete or inexact supervisions to improve model performance~\cite{Yang2021ASO,han2020semi,Mittal2021SemiSupervisedSS,zhai2022ass,Kim2021WeaklysupervisedDL}.
Han \etal adopted a generative adversarial network (GAN)-based model for semi-supervised breast tumor segmentation on BUS images~\cite{han2020semi}.
This method employed an evaluation network to assess the quality of the segmentation outcomes in order to enhance the model performance through an adversarial training strategy.
Mittal \etal proposed a dual-branch model using the consistency regularization technique, which combined a GAN-based network for segmentation and a multi-label teacher network to filter false positive segmentation predictions to improve model performance~\cite{Mittal2021SemiSupervisedSS}. 

In this study, we proposed a novel MTL method to address these two problems for joint breast tumor lesion localization and classification based on a disentangled semi-supervised learning strategy and attention mechanism. 
The proposed model was composed of a shared feature extractor appended by an auxiliary lesion-aware network and a classifier for joint lesion localization and classification.
Multiple attention modules were employed in the auxiliary lesion-aware network to optimize the multi-scale intermediate feature maps from the feature extractor.
This design can leverage the intensity-level and geometrical-level knowledge and improve the representativeness of the extracted feature maps by focusing on the lesion region via the channel and spatial attention,
thus achieving better performance in both classification and localization tasks.
The disentangled semi-supervised learning strategy was designed for training the model by use of incomplete training datasets with partial lesion location annotations.
It was adopted from the pseudo-labeling method, which is a simple but efficient semi-supervised learning approach~\cite{Lee2013PseudoLabelT}.
By assigning high-confident pseudo-labels to unlabeled images to increase the number of labeled training samples,
this learning strategy can significantly reduce the burden on data annotation and fully utilize the unlabeled data to improve localization performance.
In addition, the proposed model was modularized so that each network component can be flexibly configured and adjusted to satisfy specific objectives in various potential applications.
Experimental results on two breast ultrasound image datasets demonstrate the effectiveness of the proposed method.
The impacts of various network factors on model performance are also investigated to gain deep insights into the designed model.

The remainder of the paper is organized as follows.
Section~\ref{sec:method} describes the proposed lesion-aware classification method.
Section~\ref{sec:imple} describes the dataset and implementation details of the proposed method, and the experimental results are shown in Section~\ref{sec:result}.
The discussion and conclusion are described in Section~\ref{sec:disc} and Section~\ref{sec:conclusion}, respectively.

\section{Methods}
\label{sec:method}

\subsection{The proposed framework architecture}
\label{subsec:framework}

\begin{figure}[!ht]
\centering
\includegraphics[width=0.8\textwidth]{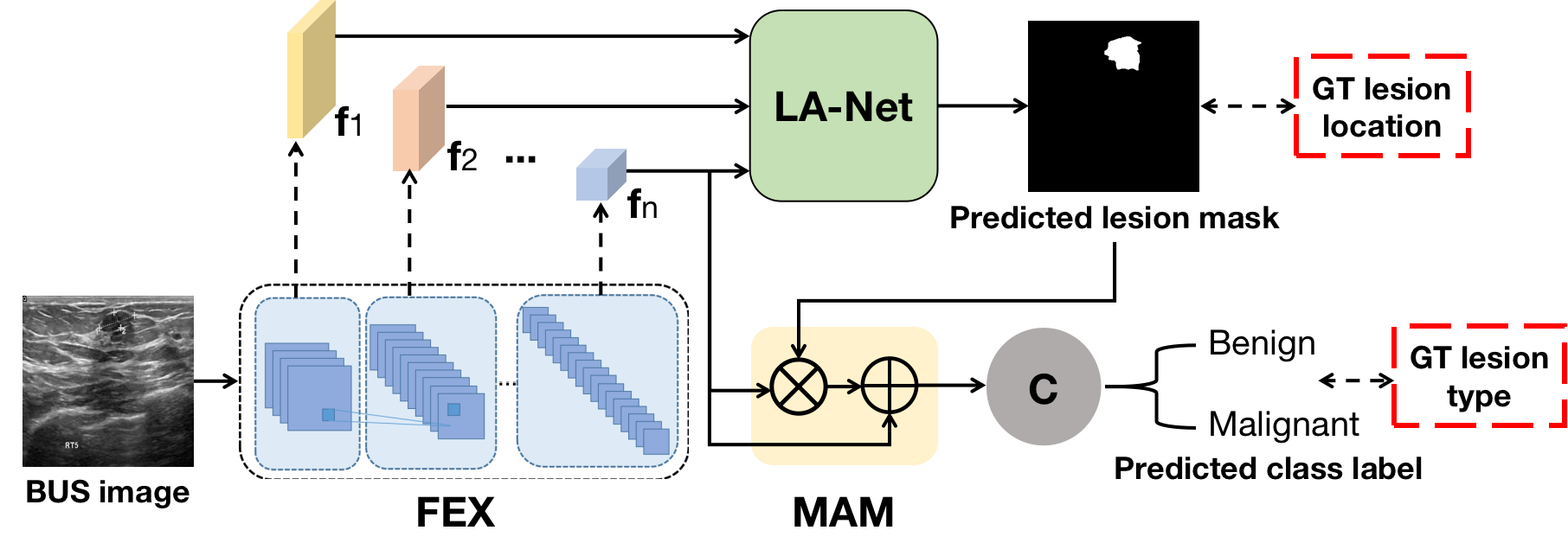}
\caption{The proposed framework for joint localization and classification. FEX: feature extractor; C: classifier; $\{\mathbf{f}_1,...,\mathbf{f}_n\}$: the extracted intermediate feature maps of FEX; MAM: mask attention module where $\bigotimes$ means element-wise multiplication and $\bigoplus$ means element-wise addition. 
}
\label{fig:model_training}
\end{figure}

As shown in Figure~\ref{fig:model_training}, the proposed framework consists of a feature extractor (FEX) followed by a classifier and an auxiliary lesion-aware network (LA-Net).
The FEX extracts feature maps from hierarchical intermediate convolutional layers, which contain rich multi-scale lesion-relevant information.
These features are shared for the classification and localization task via two branches.
The auxiliary LA-Net branch leverages these multi-scale feature maps via multiple attention modules to predict the potential lesion location. 
The classifier branch predicts the class labels by combining the learned lesion localization knowledge from LA-Net with extracted feature maps of FEX through a self-attention module.
This design explicitly utilizes correlation and alleviates the potential optimization conflicts between the classification task and localization task.
The model architecture is discussed in terms of each network as follows.

\subsubsection{Feature extractor (FEX)}
\label{subsec:featnet}

The FEX is hierarchically structured with $n$ convolutional blocks.
Given a 2-dimensional BUS image $X\in\mathbb{R}^{M\times N}$,
a set of feature maps $\{\mathbf{f}_1, ..., \mathbf{f}_n\}$ are extracted by each of the convolutional blocks, which will be used as the input of LA-Net.
Only the top extracted feature map $\mathbf{f}_n$ with the lowest dimensionality is used as the input of the classifier to improve the classification robustness.
The mathematical representation of FEX is:
\begin{equation}
\mathbf{\{f_1, f_2, ..., f_n\}} = F(X, \Theta_F),
\label{eq:f}
\end{equation}
where $F$ represents the mapping function of FEX parameterized by $\Theta_F$. 

\subsubsection{Lesion-aware network (LA-Net)}
\label{subsec:la-net}
\begin{figure}[!ht]
    \centering
    \includegraphics[width=0.6\textwidth]{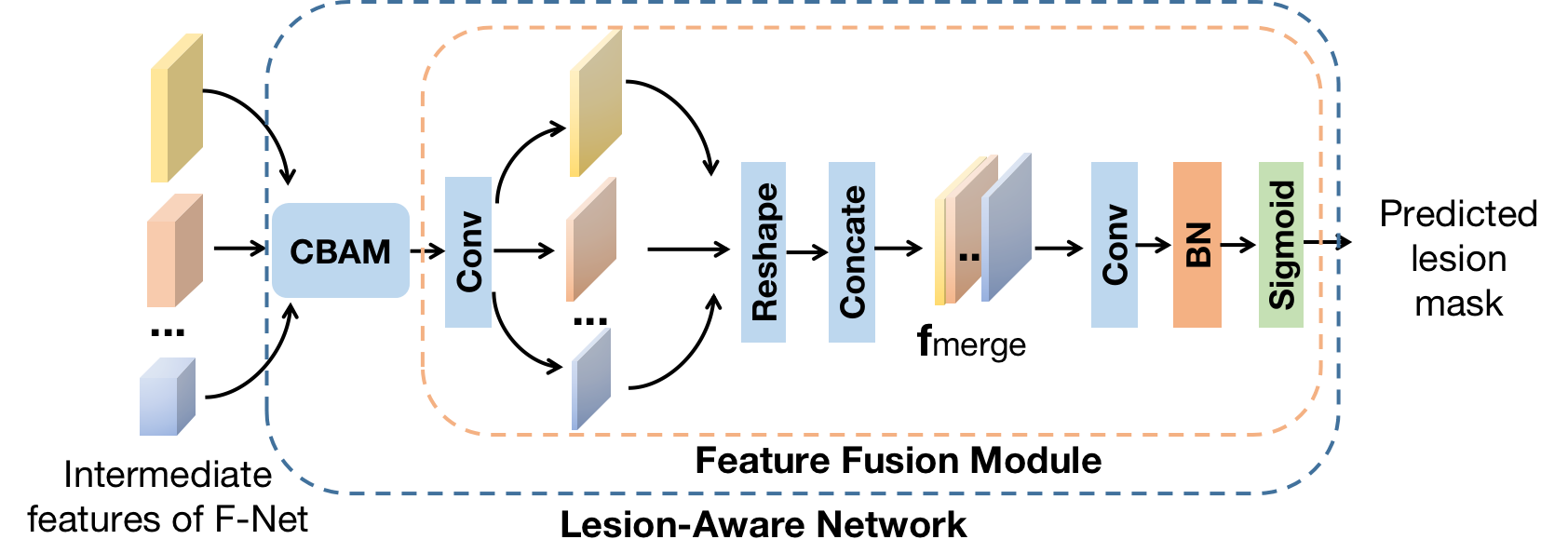}
    \includegraphics[width=0.6\textwidth]{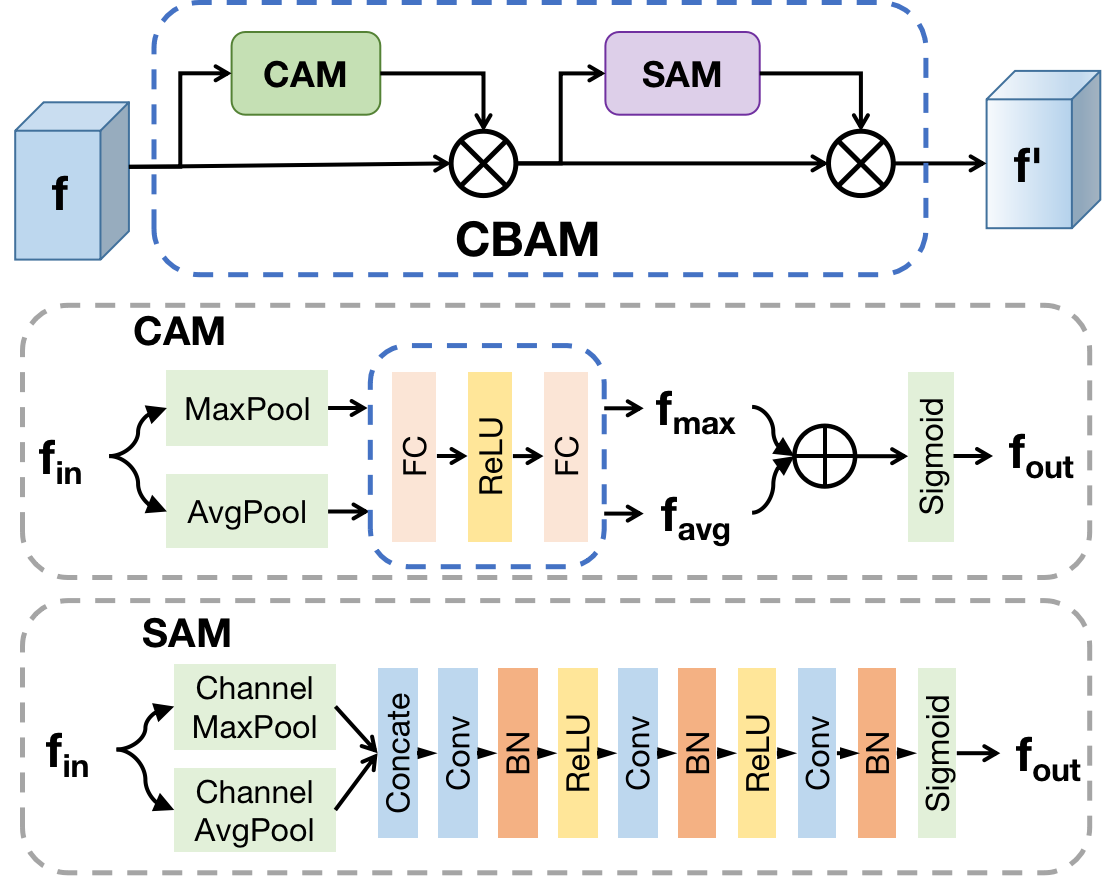}
    \caption{Network architecture of lesion-aware network (LA-Net). CAM: channel attention module; SAM: spatial attention module. 
    In the CAM module, MaxPool means applying maximum pooling of a feature map on the width and height axes to output a vector. AvgPool represents average pooling applied in the same way. The $\bigotimes$ and $\bigoplus$ denote element-wise multiplication and addition, respectively. 
    In the SAM module, Channel MaxPool means applying maximum pooling of a multi-channel feature map on the channel axis to output a single-channel feature map. Channel AvgPool applies average pooling in the same way.}
    \label{fig:la-net}
\end{figure}

The architecture of LA-Net is shown in Figure~\ref{fig:la-net}. 
A shared convolutional block attention module (CBAM)~\cite{woo2018cbam} is employed to process the set of extracted multi-scale intermediate features $\{\mathbf{f}_1,...,\mathbf{f}_n\}$ from FEX.
CBAM includes a channel attention module (CAM) and a spatial attention module (SAM).
By fusing channel attention and spatial attention, CBAM can exploit channel and spatial knowledge to enhance the informativeness of extracted feature maps.
Next, a feature fusion module is designed to fuse the CBAM-processed feature maps and predict the lesion location mask.
First, a convolutional (Conv) layer is employed to squeeze the multiple channels of the input feature map into a single channel, distilling the learned knowledge to highlight the lesion regions of interest (ROIs).
These squeezed feature maps are resized to the size of feature $\mathbf{f}_n$ and concatenated into a feature map $\mathbf{f}_{merge}$ with dimension of $S_n\times S_n\times n$, where $S_n$ is the width and height of $\mathbf{f}_n$.
The merged feature map is processed by a Conv layer, batch normalization (BN) layer and sigmoid activation layer to output a lesion location mask $\mathbf{Y}_{pixel}$ with the shape of $S_n\times S_n$.
Each pixel on the lesion location mask indicates its probability of belonging to the lesion or background, 
\begin{equation}
    \mathbf{P}(\mathbf{Y}_{pixel}|X) = D(\{\mathbf{f}_1, ..\mathbf{f}_n\}, \Theta_{D}),
    \label{eq:d}
\end{equation}
where $D$ represents the mapping function of LA-Net which is parameterized by trainable parameters $\Theta_{D}$.

\subsubsection{Classifier}
The last extracted feature map $\mathbf{f}_{n}$ of FEX is first enhanced by $\mathbf{P}(\mathbf{Y}_{pixel})$ through a mask-attention module (MAM), as shown in Figure~\ref{fig:model_training}.
Then the classifier uses the enhanced feature map $\mathbf{f}_{att}$ to predict the probability of the input BUS image belonging to each of $K$ classes, $P(\mathbf{Y}_{cls}|X)$.
With the MAM design, the lesion location information is introduced into the classification process, which improves the classification performance by learning features to focus more on the explicit ROIs without the disturbance of the noisy background.
The classification process can be described as:
\begin{align}
    \mathbf{f}_{att} = \mathbf{f}_n + \mathbf{f}_n \otimes \mathbf{P}(\mathbf{Y}_{pixel}), \\
    \mathbf{P}(\mathbf{Y}_{cls}|X) =C(\mathbf{f}_{att},\Theta_{C}),
    \label{eq:c}
\end{align}
where $C$ is the mapping function of the classifier parameterized by $\Theta_C$ and $\bigotimes$ denotes element-wise multiplication.

\subsection{Disentangle semi-supervised learning strategy}
\label{subsec:train} 

A disentangle semi-supervised training strategy was proposed to train this framework using training datasets with fully-annotated class labels but incomplete location labels.
As shown in Algorithm~\ref{algo:algorithm},
the training process consists of two stages. 
The first stage is to train the LA-Net by use of data with localization annotations.
The FEX and LA-Net are jointly trained via the minimization of the $L_{loc}$ loss as follows:
\begin{equation}
    \label{eq:loc-loss}
    L_{loc} = \mathbb{E}[-\mathbf{\bar{Y}}^T_{pixel} log(\mathbf{P}_{pixel})] 
\end{equation}
where $L_{loc}$ measures the geometric difference between GT lesion location mask and predicted lesion location mask,
$\mathbf{P}_{pixel}$ represents the probabilities of pixels on the predicted location mask belonging to the lesion, and $\mathbf{\bar{Y}}_{pixel}$ represents the GT lesion location mask.

The second stage is to further optimize the model parameters by use of the whole dataset in a semi-supervised learning fashion.
The parameters of the FEX, LA-Net, and classifier are optimized by minimizing a hybrid loss $L_{hyb}$ as below.
\begin{equation}
    \label{eq:hyb}
    \begin{aligned}
    L_{hyb} &= \lambda L_{cls}(\Theta_F,\Theta_C,\Theta_{D}) + (1-\lambda) L_{loc}'(\Theta_{F}, \Theta_D), \\ 
    L_{cls} &= \mathbb{E}[-\mathbf{\bar{Y}}^T_{cls} log(\mathbf{P}_{cls})], \\
    L_{loc}'&= \mathbb{E}[-\mathbf{\bar{Y}}^T_{pixel} log(\mathbf{P}_{pixel})] + \alpha \mathbb{E}[-\mathbf{\bar{Y'}}^T_{pixel} log(\mathbf{P'}_{pixel})], \\
    \mathbf{\bar{Y}'}_{pixel} &=binarize(\mathbf{P'}_{pixel}, \tau), \\
    \end{aligned}
\end{equation}
where $L_{cls}$ measures the discrepancy between the ground-truth (GT) class labels and predicted class labels through classifier $C$;
$\mathbf{P}(\mathbf{Y}_{cls})$ represents the probabilities of input data $X$ belonging to each of $K$ class and $\mathbf{\bar{Y}}_{cls}$ is an one-hot GT label vector with $K$ elements;
$L_{loc}'$ represents the localization loss using the whole dataset;
$\lambda \in[0,1]$ is the weighting factor to balance the contribution of $L_{loc}$ and $L_{cls}$;
$\Theta_F$, $\Theta_C$, and $\Theta_{D}$ represent the trainable parameters of network FEX, classifier, and LA-Net, respectively.
Particularly, $L_{loc}'$ consists of two loss terms from both labeled and unlabeled data.
The first loss term represents the measurement result from the labeled data.
The second loss term corresponds to the unlabeled data,
while $\mathbf{P}'_{pixel}$ represents the probabilities of predicted location labels from unlabeled data, and $\mathbf{\bar{Y}'}_{pixel}$ is the pseudo-labeling lesion location label which is determined by a binarize function with prediction confidence threshold $\tau$.
Parameter $\alpha$ is the coefficient to balance the loss terms from labeled and unlabeled data.
In this study, cross-entropy was used for both $L_{cls}$, $L_{loc}$ and $L_{loc}'$.

Noticeably, pre-training LA-Net in the first stage can benefit the training convergence and stability of the model in the second stage.
In the second stage, when the training sample $X$ doesn't have the annotated lesion location, $L_{loc}$ is penalized by coefficient $\alpha$ to reduce the disturbance of inaccurate pseudo-GT labels.
Therefore, the optimization of the LA-Net in the second stage is mainly driven by the classification loss, making the lesion localization information acquired by LA-Net more suitable for classification.
The classification performance can be improved even on datasets with partially-annotated location labels,
greatly reducing the burden of data annotation.

\begin{algorithm}
\SetAlgoVlined
\DontPrintSemicolon
\SetKw{param}{Networks with Paramters:}
\SetKw{hyper}{Training Hyperparameters:}
\SetKw{start}{Start Training}
\SetKw{stop}{End Training}
\caption{Minibatch training of the proposed framework}
\KwIn{Training dataset $\mathcal{X}$ with $N$ pairs of image and GT class label; Within them, there is a sub-dataset $\mathcal{X}_l\in \mathcal{X}$ with $M$ images annotated with lesion location labels ($M<N$)}
\param{\textnormal{FEX with $\Theta_F$; classifier with $\Theta_C$; LA-Net with $\Theta_{D}$}}

\hyper{\textnormal{Minibatch size: $m$; the number of lesion location pre-training epochs: $t_l$; the number of classifier training epochs: $t_c$;}}

\vspace{0.25cm}
\start

Stage 1: Pre-train the LA-Net with full location supervision $\mathcal{X}_l$

Current training iteration: $i=1$\;

\While{$i\le t_l$}{
    The number of trained images in the current epoch: $n$; Set $n=0$\;
    \While{$n\le M$}{
        \begin{minipage}{0.85\linewidth}
        \begin{enumerate}
        \item Sample minibatch of $m$ images $\{X_l^{(1)},X_l^{(2)},...,X_l^{(m)}\}$ from input data and their corresponding lesion location labels $\{\bar{Y}_{pixel}^{(1)},\bar{Y}_{pixel}^{(2)},...,\bar{Y}_{pixel}^{(m)}\}$\; 
        \item Forward the minibatch data through the FEX and LA-Net and output mask prediction $\hat{Y}_{pixel}$\;
        \item Compute the $L_{loc}$ as Equation.~\ref{eq:loc-loss}\; 
        \item Update the FEX and LA-Net by ascending its stochastic gradient: $\bigtriangledown_{\Theta_F,\Theta_{D}}L_{loc}$\;
        \item n=n+m\;
        \end{enumerate}
        \end{minipage}
    }
    Iteration $i=i+1$\;
}

\KwOut{Trained LA-Net with parameters $\bar{\Theta}_{D}$}

\vspace{0.25cm}
Stage 2: Train three networks with incomplete dataset $\mathcal{X}$ in a semi-supervised learning strategy

\start

Load LA-Net with trained parameters $\bar{\Theta}_{D}$ obtained in Stage 1\;
Current training iteration: $i=1$\;

\While{$i\le t_c$}{
    The number of trained images in the current epoch: $n$; Set $n=0$\;
    \While{$n\le N$}{
        \begin{minipage}{0.85\linewidth}
        \begin{enumerate}
        \item Sample minibatch of $m$ labeled images $\{X^{(1)},X^{(2)},...,X^{(m)}\}$ from input data and their corresponding class labels $\{\bar{Y}_{cls}^{(1)},\bar{Y}_{cls}^{(2)},...,\bar{Y}_{cls}^{(m)}\}$\ and location labels $\{\bar{Y}_{pixel}^{(1)},\bar{Y}_{pixel}^{(2)},...,\bar{Y}_{pixel}^{(m)}\}$; similarly, sample minibatch of $m'$ unlabeled images $\{X'^{(1)},X'^{(2)},...,X'^{(m')}\}$ with class labels $\{\bar{Y'}_{cls}^{(1)},\bar{Y'}_{cls}^{(2)},...,\bar{Y'}_{cls}^{(m')}\}$ but without location labels\;
        \item Forward the minibatch data through the FEX, classifier and LA-Net and output the predicted class label $\hat{Y}_{cls}$ and predicted lesion location mask $\hat{Y}_{pixel}$ from the labeled data and $\hat{Y'}_{cls}$ and $\hat{Y'}_{pixel}$ from unlabeled data\;
        \item Compute the $L_{hyb}$ as Equation.~\ref{eq:hyb}\;
        \item Update the FEX, classifier and LA-Net by ascending its stochastic gradient: $\bigtriangledown_{\Theta_F, \Theta_C, \Theta_{D}}L_{hyb}$\;
        \item n=n+m+m'\;
        \end{enumerate}
        \end{minipage}
    }
    Iteration $i=i+1$\;
}
\stop 

\KwOut{Trained model weights of the FEX, classifier and LA-Net} 
\label{algo:algorithm}
\end{algorithm}

\section{DATASET \& METHOD IMPLEMENTATION}
\label{sec:imple}

\subsection{Dataset}
\label{subsec:dataset}

\begin{figure}[!ht]
\centering
\includegraphics[width=0.6\textwidth]{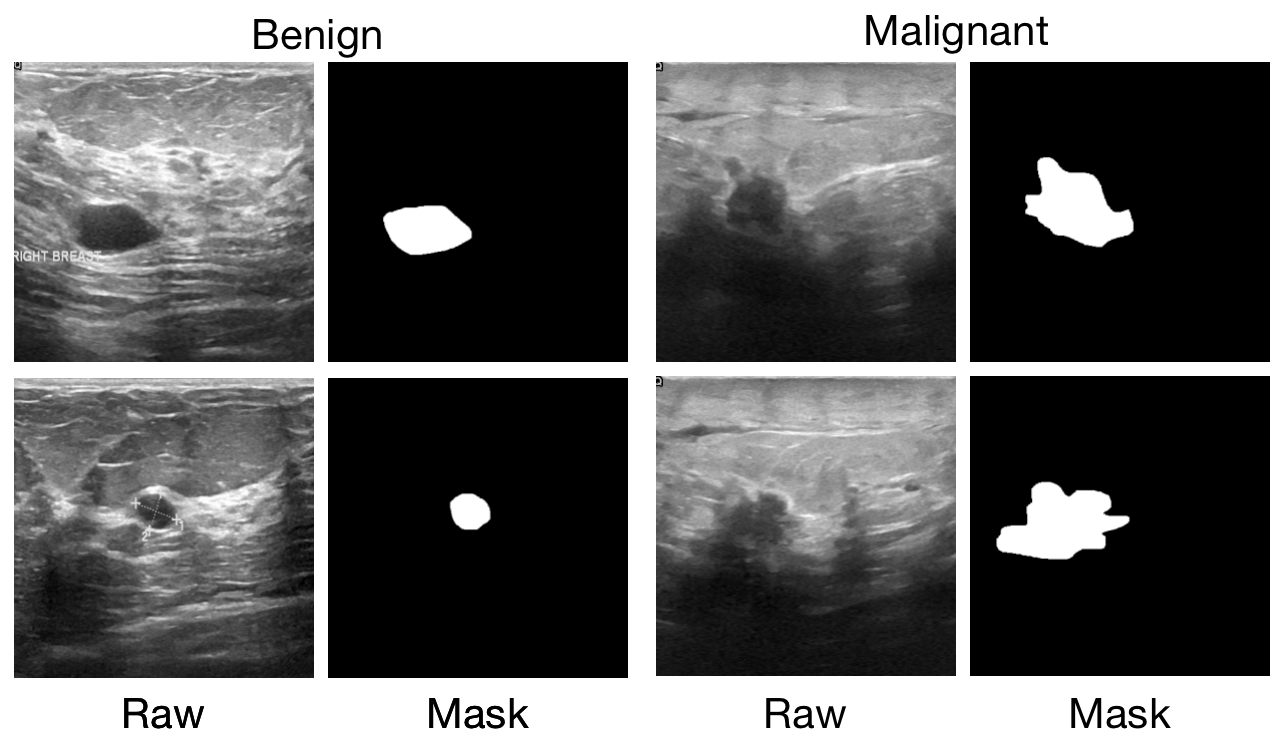}
\caption{Example images of BUD dataset. The left two columns show the ultrasound images with benign tumors and corresponding segmentation masks, respectively. The right two columns show the malignant tumor examples and the lesion masks.
}
\label{fig:us_bud}
\vspace{0.1in}
\end{figure}

The proposed method was evaluated on two breast ultrasound image datasets. The first dataset is a public dataset, the Breast Ultrasound Dataset (BUD)~\cite{AlDhabyani2020DatasetOB}.
This database includes a total of 780 images with image size of $500\times 500$ pixels after scanning 600 female patients acquired by LOGIQ E9 ultrasound system and LOGIQ E9 Agile ultrasound system.
This dataset was fully-annotated with segmentation masks as the lesion localization labels.
The details of this dataset are shown in the table~\ref{tab:bud_dataset}, and several example images are shown in Figure~\ref{fig:us_bud}.

\begin{figure}[h]
    \centering
    \includegraphics[width=0.6\textwidth]{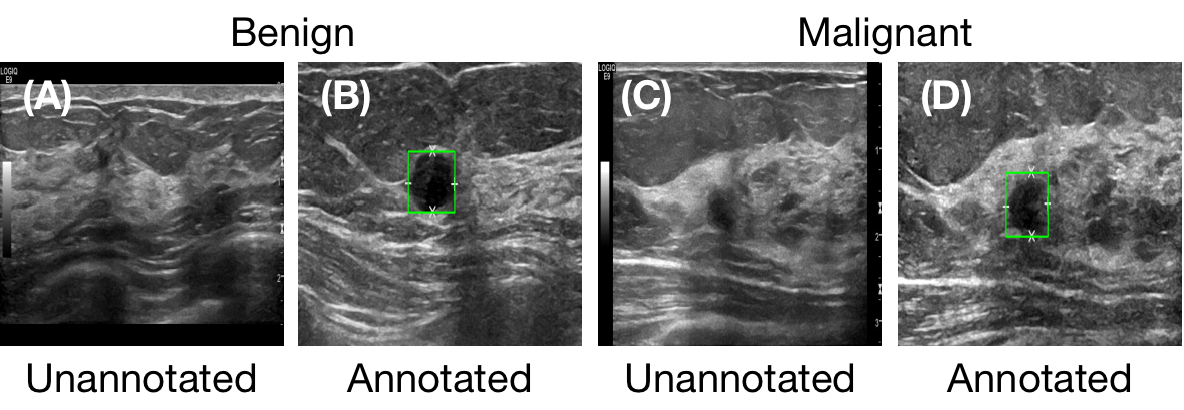}
    \caption{Examples of F-MBUD dataset. Image (A) and (B) show the benign examples annotated without/with the lesion location (shown in the green bounding box), respectively. Image (C) and (D) correspond to the malignant tumor examples.
    }
    \label{fig:us_mayo}
    \vspace{0.1in}
\end{figure}

The second dataset includes breast tumor ultrasound images collected at Mayo Clinic (MBUD).
A total of 160 scanning videos with image size of $854\times 500$ pixels were collected after scanning 136 patients with LOGIQ E9 ultrasound system.
A total of 22202 images were extracted from these videos and annotated with class labels.
Within them, there were 384 images manually annotated with bounding boxes as the lesion localization labels.
Several example images are shown in Figure~\ref{fig:us_mayo}.
We named the dataset with whole images but partial location annotations as P-MBUD, and the sub-dataset with 384 fully-annotated images as the F-MBUD. 
The details of the F-MBUD and P-MBUD datasets are shown in Table~\ref{tab:mayo_dataset}.
Particularly, P-MBUD was used as the incomplete dataset to evaluate the performance of our proposed semi-supervised learning strategy.

\begin{table}[!ht]
    \centering
    \begin{threeparttable}
    \begin{tabular}{c c c}
    \hhline{===}
    \textbf{Class label} & \textbf{Training images} & \textbf{Testing images} \\
    \hline
    Benign & 397 &   40 \\
    Malignant & 170 &  40 \\ 
    \hhline{===}
    \end{tabular}
    \captionof{table}{Details of the BUD dataset. }
    \label{tab:bud_dataset}
    \end{threeparttable}
    \vspace{0.1in}
\end{table}

\begin{table}[!ht]
\centering
\begin{threeparttable}
\begin{tabular}{c c c}
\hhline{===}
\textbf{Class label} & \textbf{Training images} & \textbf{Testing images} \\
\hline
Benign & 13440 (220) & 2435 (36)  \\
Malignant & 4954 (108) &  1373 (20)\\
\hhline{===}
\end{tabular}
\captionof{table}{Details of the P-MBUD dataset and F-MBUD dataset (shown in parentheses).}
\label{tab:mayo_dataset}
\end{threeparttable}
\vspace{0.1in}
\end{table}

\subsection{Network architecture settings}
\label{subsec:net_arch_set}
FEX was designed as the standard residual neural networks (ResNets)~\cite{He2016DeepRL}, which is a powerful CNN network architecture in BUS image classification tasks~\cite{ding2022joint,mishra2021breast}.
ResNets consist of multiple residual blocks which apply shortcut connections between non-adjacent convolutional layers to address gradient vanishing issue and improve training stability.
The extracted intermediate features $\{\mathbf{f}_1,...,\mathbf{f}_n\}$ have dimensions of $\{(C_1, S_1, S_1),...,(C_n, S_n, S_n)\}$, respectively, where $C_i$ represents the number of channels of the $i$th extracted feature map, $S_i$ represents the $i$th extracted feature map size and $i\in[1,n]$.
In this study, ResNet18 and ResNet50 were used as FEX to evaluate the effect of network depths.
Four feature maps ($n=4$) were extracted from the intermediate residual blocks of FEX, which down-sampled the feature map size by half.

The architecture of LA-Net is shown in Figure~\ref{fig:la-net}.
CBAM was shared by each extracted feature map from FEX, where the original CBAM network settings were employed here~\cite{woo2018cbam}.
Given an input multi-channel feature map $\mathbf{f}_i$, CBAM refined this feature via CAM and SAM and output a distilled feature map.
The feature fusion module was designed to integrate the multi-scale CBAM-processed feature maps.
A Conv layer was first employed to squeeze the channels of each CBAM-processed feature map into a single channel.
Then all four feature maps were resized to the size of the last extracted feature $\mathbf{f}_4$ and concatenated into a merged feature map $\mathbf{f}_{merge}$.
Finally, a Conv layer, a batch normalization (BN) layer and a sigmoid activation layer were employed to predict the lesion location mask by use of $\mathbf{f}_{merge}$. 

The classifier was added on top of the FEX, consisting of one average pooling layer, one FC layer and a softmax activation layer. 
The FC layer had $K$ neurons, which corresponded to the number of classes and $K=3$ in this study.

\subsection{Settings for framework training and testing}
\label{subsec:train_param}
The training images described in Table.~\ref{tab:bud_dataset} and Table.~\ref{tab:mayo_dataset} were randomly divided into the training dataset (85\%) and validation dataset (15\%).
The ResNet pre-trained on ImageNet dataset~\cite{Deng2009ImageNetAL} was used to initialize the parameters of FEX.
The model parameters of the LA-Net were initialized using Xavier initialization scheme~\cite{pmlr-v9-glorot10a}.
In each training iteration, a mini-batch of 8 labeled and unlabeled images were sampled randomly from the training dataset.
The input images were first preprocessed with the following data augmentation techniques, including
random rotation, random horizontal flip, random vertical flip and random color jittering.
The preprocessed images were resized to $M\times M$ pixels by bi-linear interpolation.
Here $M\in\{256, 512\}$ in this study to investigate the impact of image size on the proposed method.
Before input into FEX, the resized images were normalized to $[0, 1]$ pixel value range.
To update the model parameters, the loss functions defined in Section~\ref{subsec:train} were calculated.
The binarization confidence threshold $\tau$ to obtain the pseudo-labeling lesion location label was 0.8. 
The weighing coefficients $\alpha$ and $\lambda$ in Equation.~\ref{eq:hyb} was set to 0.1 and 0.5, respectively.
With loss calculated, the Adam stochastic gradient algorithm~\cite{Kingma2015AdamAM} was employed as the optimization algorithm with an initial learning rate $lr=0.001$.
The framework was trained on the training dataset for 100 epochs.

After training, the model with the highest accuracy on the validation dataset was further tested on the testing dataset.
To assess the network stability, the procedure described above was repeated three times.
Mean values and standard deviations of the results were calculated to evaluate the performance and stability of the trained model.
Particularly, when the location labels were in the form of bounding boxes, the predicted lesion mask was converted to a bounding box by choosing the minimum rectangle that held the predicted lesion region before evaluation.
The proposed framework was implemented by use of PyTorch 1.7.0.
The training and validation process was performed using Nvidia GeForce GTX 1080ti GPUs.

\subsection{Other methods for comparison}
\label{subsec:comp}
In this study, a typical and effective MTL method introduced by Rasaee \etal was employed as comparison~\cite{rasaee2021explainable}.
The classification network was a ResNet50~\cite{He2016DeepRL}. 
The feature map extracted from the top Conv layer of the classification network was used as the input of the segmentation network.
The segmentation network had three upsampling blocks, where each block consisted of a deconvolutional layer, a BN layer and a ReLU activation layer.
These two networks were updated simultaneously by minimizing a hybrid loss consisting of the localization loss and classification loss.
This MTL method was called RMTL in the rest of the paper.
RMTL method requires fully-segmented datasets with both segmentation masks and class labels.
Therefore, the RMTL method was trained and tested on BUD and F-MBUD datasets following the same settings discussed in Section~\ref{subsec:train_param}.

\subsection{Performance evaluation metrics}
\label{subsec:metric}
In this study, since LA-Net could provide lesion location prediction, both the classification and the localization performances of the framework were evaluated. 
Precision, specificity, sensitivity, and F1-score (or dice coefficient),
were employed as the metrics to evaluate the classification performance.
To evaluate the localization performance, Jaccard Similarity Index (JSI) was employed, which is also known as the intersection over union (IoU). 
It is the ratio of overlap area with respect to the union area between the predicted segmentation contour or bounding box and the GT segmentation contour or bounding box of a class. 

\section{Experimental Results}
\label{sec:result}

\subsection{Classification performance of the proposed method}
\label{subsec:cls_pfm}
\begin{figure}[!ht]
    \centering
    \begin{subfigure}[b]{0.9\textwidth}
        \includegraphics[width=\textwidth]{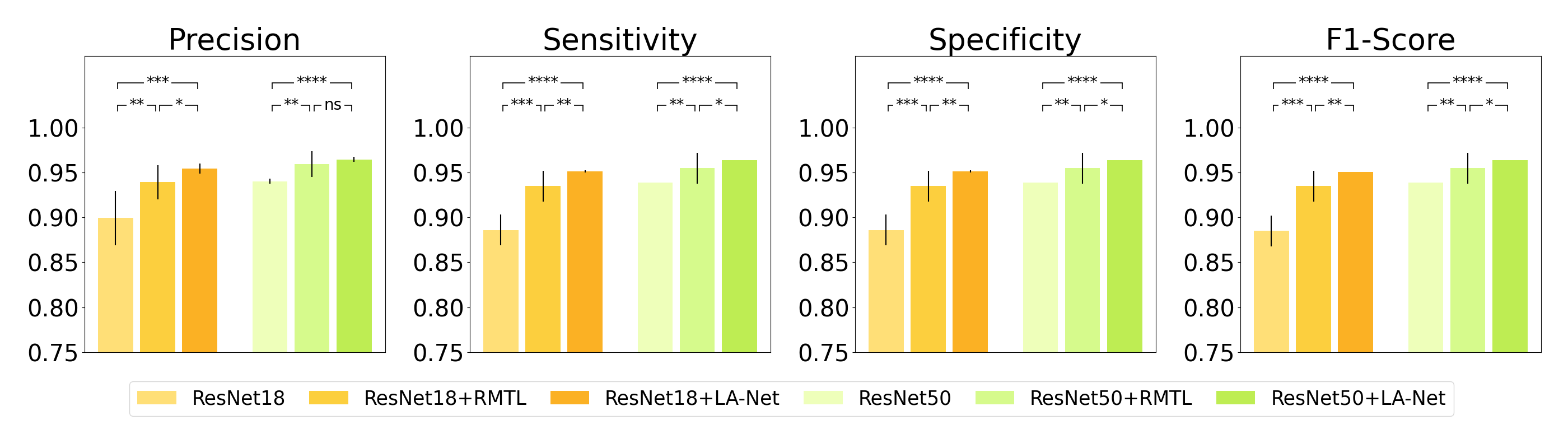}
        \caption{Image size: 256}
        \label{subfig:bud_256}
    \end{subfigure}
    \hfill
    \begin{subfigure}[b]{0.9\textwidth}
        \includegraphics[width=\textwidth]{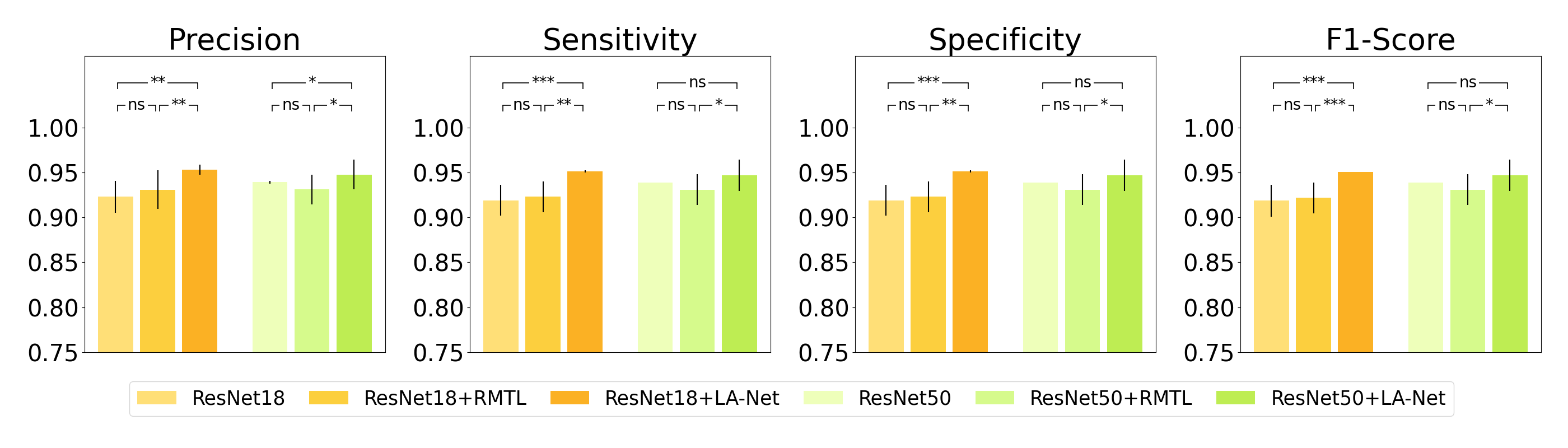}
        \caption{Image size: 512}
        \label{subfig:bud_512}
    \end{subfigure}
    \caption{Classification performance of the proposed method on BUD dataset. ResNet50 or ResNet18: Vanilla ResNet used as FEX; \textbf{+LA-Net}: our proposed method.
    The statistical significance symbol above indicates the $\it{t}$-test result of the selected pair of methods, whose null hypothesis is that the average metric of the left method is less than or equal to the right method's. Here $ns$ represents $\it{p}$-value$>0.05$ (no significant improvement). The symbols $*$, $**$, $*{*}*$, and $*{**}*$ indicate the $\it{p}$-value is less than $0.05$ and $0.01$, $0.01$, $0.001$, and $0.0001$ (with significant improvement), respectively.}
    \label{fig:net_effect_bud}
\end{figure}

The classification performance of the proposed method on the BUD dataset is shown in Figure~\ref{subfig:bud_256}.
The performance achieved by vanilla ResNets was considered as the baseline.
When using ResNet18 as FEX and setting the input image size as $256\times 256$ pixels,
our method achieved 5.6\%, 6.5\%, 6.5\%, 6.6\% improvement in terms of sensitivity, specificity, precision, and F1-score, respectively.
The unpaired $\it{t}$-test showed that our method significantly improved the classification performance compared to both the vanilla ResNet and RMTL method.
This result indicates the superior effect of LA-Net in improving classification performance.
Similar results were observed when using ResNet50 as FEX, while the improvement degree was smaller.
Considering ResNet50 can usually extract feature maps with richer information for classification, this result implies that network depth can influence the classification performance of the model and the proposed method may work better when the network is shallower. 

When the input image size was up-scaled to $512\times 512$ pixels,
our method still significantly improved the classification performance by using ResNet18 as FEX, as shown in Figure~\ref{subfig:bud_512}.
However, no significant improvement was observed when using ResNet50.
For comparison, the RMTL method achieved no improvement or even degradation using either ResNet18 or ResNet50.
Our proposed method achieved significantly better performance than the RMTL method, indicating the effectiveness of our method in alleviating the feature sharing issue and improving model performance.

\begin{figure}[!ht]
    \centering
    \begin{subfigure}[b]{0.9\textwidth}
        \includegraphics[width=\textwidth]{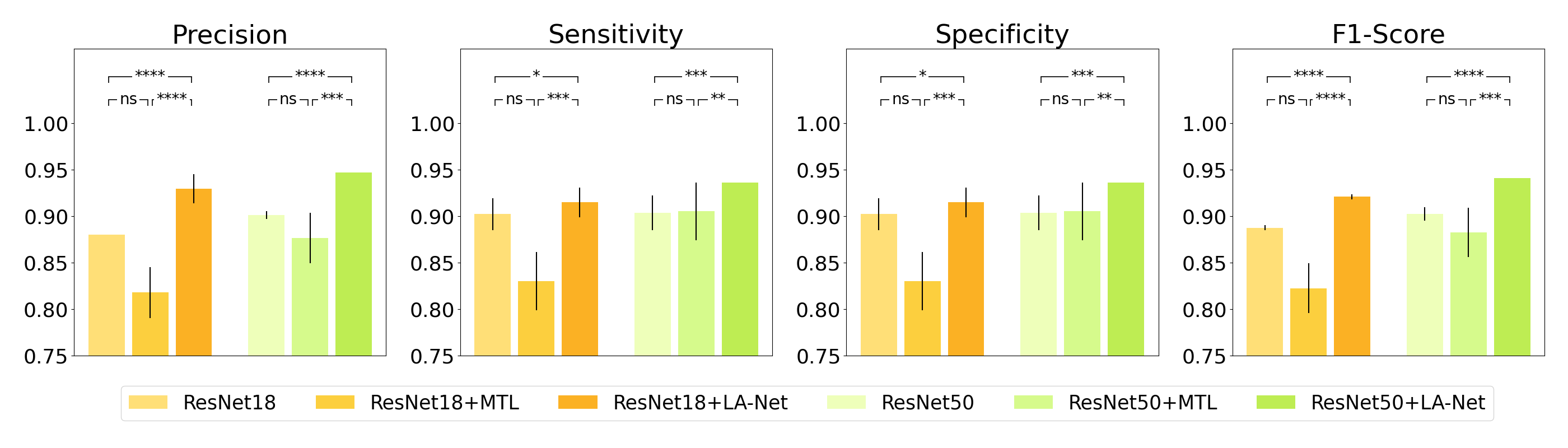}
        \caption{Image size: 256}
        \label{subfig:mayo_256}
    \end{subfigure}
    \hfill
    \begin{subfigure}[b]{0.9\textwidth}
        \includegraphics[width=\textwidth]{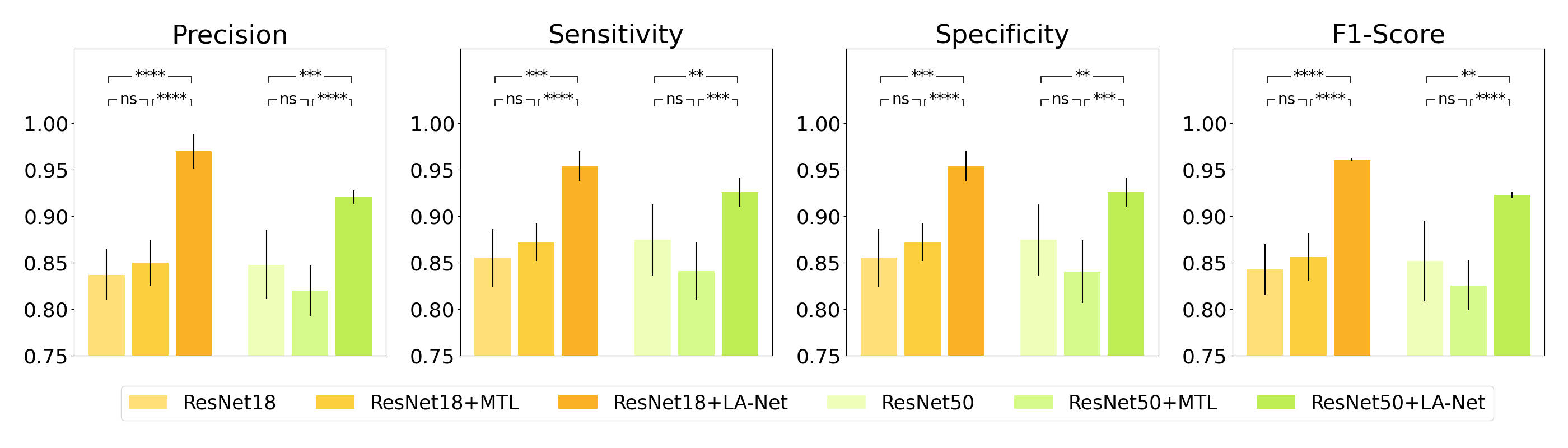}
        \caption{Image size: 512}
        \label{subfig:mayo_512}
    \end{subfigure}
    \caption{Classification performance of the proposed method on the F-MBUD dataset. The meanings of symbols are identical to those used in Figure~\ref{fig:net_effect_bud}}
    \label{fig:net_effect_mayo}
\end{figure}

As shown in Figure~\ref{fig:net_effect_mayo}, similar results were observed when using our proposed method on the F-MBUD dataset.
However, the RMTL method achieved degraded performance.
This is probably because the RMTL method requires a complete and accurate segmented dataset, while the F-MBUD dataset only provides lesion bounding boxes.
The inaccurate supervision exacerbates the issue of feature sharing conflict, leading to degraded classification performance.

\subsection{Localization performance of the proposed method}
\label{subsec:loc_pfm}

\begin{figure}[!ht]
    \centering
    \includegraphics[width=0.7\textwidth]{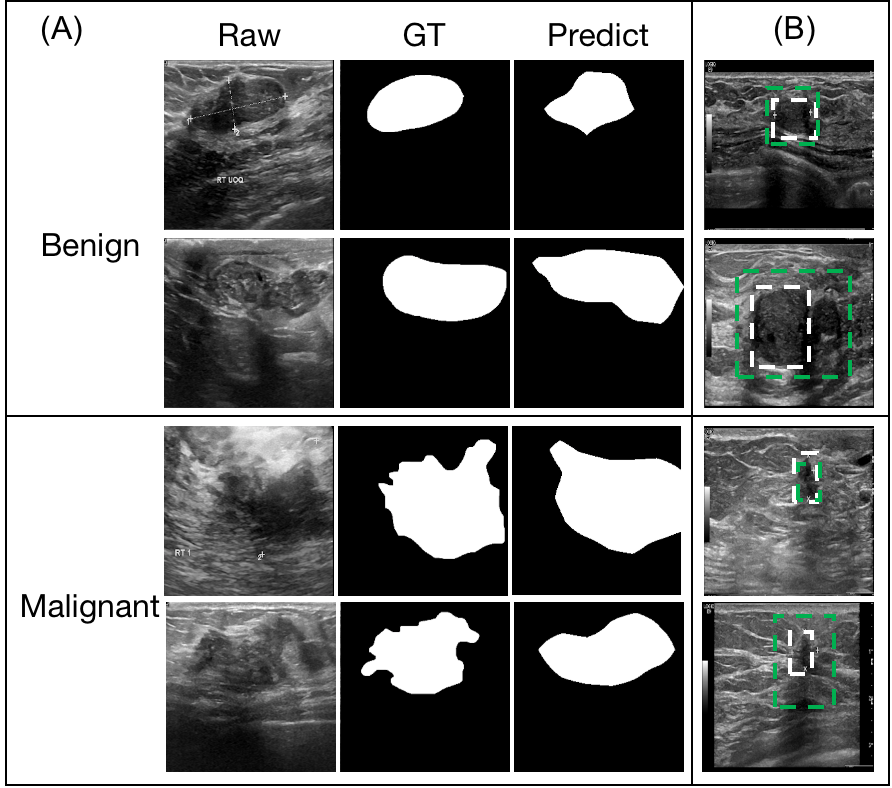}
    \caption{Predicted lesion masks with the proposed method on (A) BUD dataset and (B) F-MBUD dataset. The white boxes show the GT lesion bounding boxes, while the green boxes show the predicted lesion locations.}
    \label{fig:mask_pred}
\end{figure}

Several BUS examples and their predicted lesion masks are shown in Figure~\ref{fig:mask_pred} for qualitative analysis of the localization performance of our method.
The lesion masks were achieved by up-sampling the output of the LA-Net to the size of the input image and then binarized with a threshold of $0.5$.
These predicted lesion locations generally show impressive overlap with their GT lesion locations, especially on the BUD dataset annotated with accurate segmentation contours.
The predicted lesion locations are less precise when using F-MBUD dataset, probably due to the location annotation in the form of bounding boxes.

The quantitative localization performance analysis was performed by measuring the JSI metric described in Section~\ref{subsec:metric}.
As shown in Table~\ref{tab:iou_busi}, when using ResNet18 as FEX, our proposed method based on the BUD dataset achieved $2.8\%$ and $8.4\%$ of JSI higher than the RMTL method by use of image size as $256\times 256$ and $512\times512$ pixels, respectively.
A similar significant improvement was observed when using ResNet50 as FEX. 
These results demonstrate the superior performance of the proposed method in lesion localization compared to the RMTL method.
The experiment result based on the F-MBUD dataset is shown in Table~\ref{tab:iou_mayo}. Larger performance improvement was observed. 
These results indicate the effectiveness of our method in improving localization performance, even on a weakly-annotated dataset.
In addition, using deeper network depth or larger input image size achieved better localization performance, implying the localization performance is influenced by various network factors as well.

\begin{table}[!ht]
    \centering
    \begin{tabular}{c | c | c}
        \hline
        Network & Image size & JSI \\
        \hline
        \multirow{2}{*}{ResNet18+RMTL} & 256 & $0.382\pm 0.039$ \\
        & 512 & $0.494\pm 0.032$ \\
        \hline
        \multirow{2}{*}{ResNet18+LA-Net} & 256 & $0.410\pm 0.007$ \\
        & 512 & $0.578\pm 0.015$  \\
        \hline
        \multirow{2}{*}{ResNet50+RMTL} & 256 & $0.406\pm 0.021$ \\
        & 512 & $0.458\pm 0.037$ \\
        \hline
        \multirow{2}{*}{ResNet50+LA-Net} & 256 & $0.425\pm 0.004$ \\
        & 512 & $0.544\pm 0.020$  \\
        \hline 
    \end{tabular}
    \caption{Localization performance of the proposed method on BUD dataset. The values are represented as mean value $\pm 95\%$ confidence interval.}
    \label{tab:iou_busi}
\end{table}

\begin{table}[!ht]
    \centering
    \begin{tabular}{c | c | c}
        \hline
        Network & Image size & JSI \\
        \hline
        \multirow{2}{*}{ResNet18+RMTL} & 256 & $0.605\pm 0.041$ \\
        & 512 & $0.646\pm 0.034$ \\
        \hline
        \multirow{2}{*}{ResNet18+LA-Net} & 256 & $0.655\pm 0.047$ \\
        & 512 & $0.739\pm 0.020$  \\
        \hline
        \multirow{2}{*}{ResNet50+RMTL} & 256 & $0.614\pm 0.034$ \\
        & 512 & $0.575\pm 0.017$ \\
        \hline
        \multirow{2}{*}{ResNet50+LA-Net} & 256 & $0.661\pm 0.036$ \\
        & 512 & $0.683\pm 0.017$  \\
        \hline 
    \end{tabular}
    \caption{Localization performance of the proposed method on F-MBUD dataset. The values are represented as mean value $\pm 95\%$ confidence interval.}
    \label{tab:iou_mayo}
\end{table}

\subsection{Ablation study to understand the effects of attention modules}
\label{subsec:abl}
An ablation study was conducted to investigate the effects of three attention modules in the proposed model, including CAM, SAM and MAM modules.
The experiment was implemented by removing each attention module individually and retraining the network with the same settings described in Section~\ref{subsec:train}.

\begin{figure}[!ht]
    \centering
    \begin{subfigure}[b]{0.9\textwidth}
        \includegraphics[width=\textwidth]{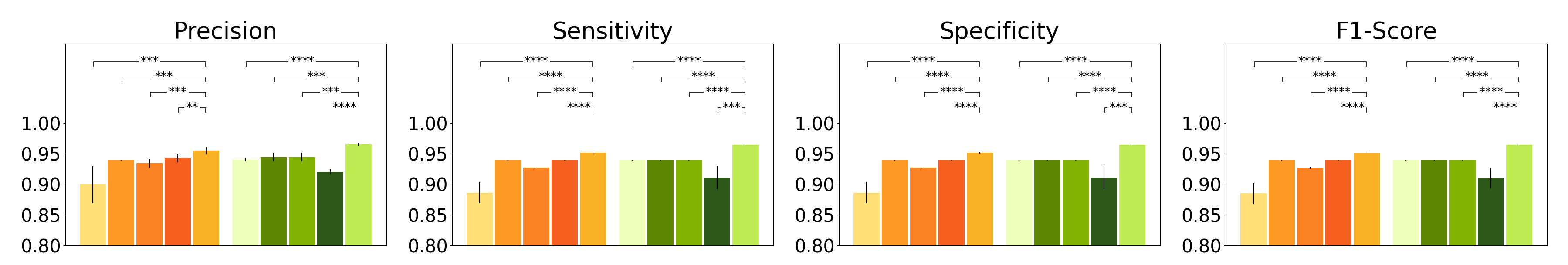}
        \caption{Image size: 256}
        \label{subfig:abl_busi_256}
    \end{subfigure}
    \hfill
    \begin{subfigure}[b]{0.9\textwidth}
        \includegraphics[width=\textwidth]{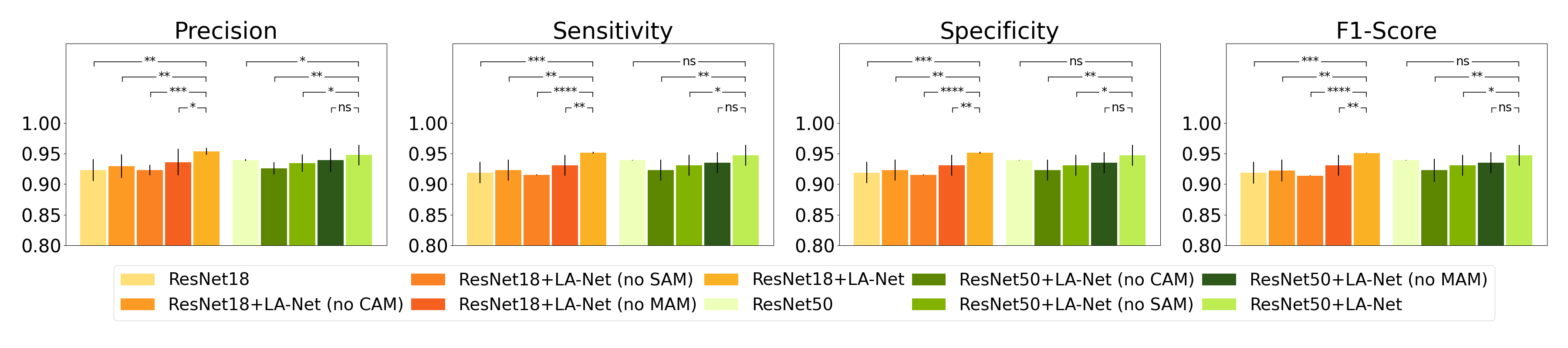}
        \caption{Image size: 512}
        \label{subfig:abl_busi_512}
    \end{subfigure}
    \caption{Ablation study to investigate the effect of attention modules in the proposed method on the BUD dataset. \textbf{no CAM}: use LA-Net without CAM module; \textbf{no SAM}: use LA-Net without SAM module; \textbf{no MAM}: use classifier without the mask attention module.}
    \label{fig:abl_busi}
\end{figure}

\begin{figure}[!ht]
    \centering
    \begin{subfigure}[b]{0.9\textwidth}
        \includegraphics[width=\textwidth]{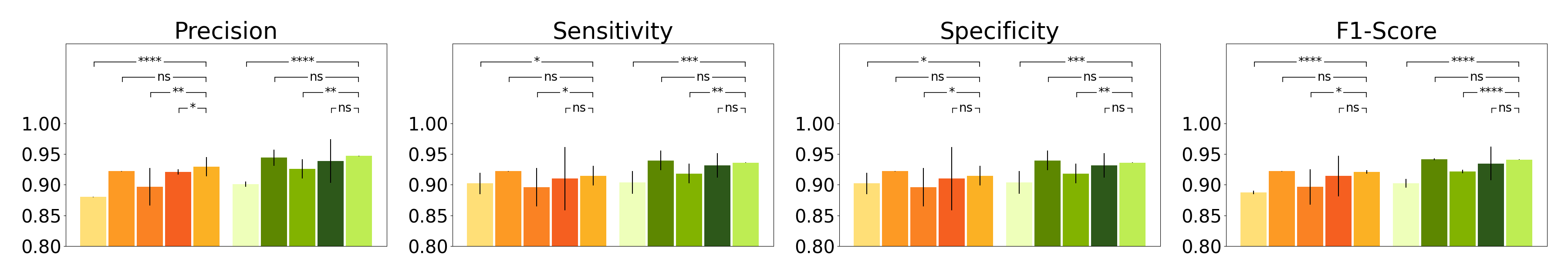}
        \caption{Image size: 256}
        \label{subfig:abl_mayo_256}
    \end{subfigure}
    \hfill
    \begin{subfigure}[b]{0.9\textwidth}
        \includegraphics[width=\textwidth]{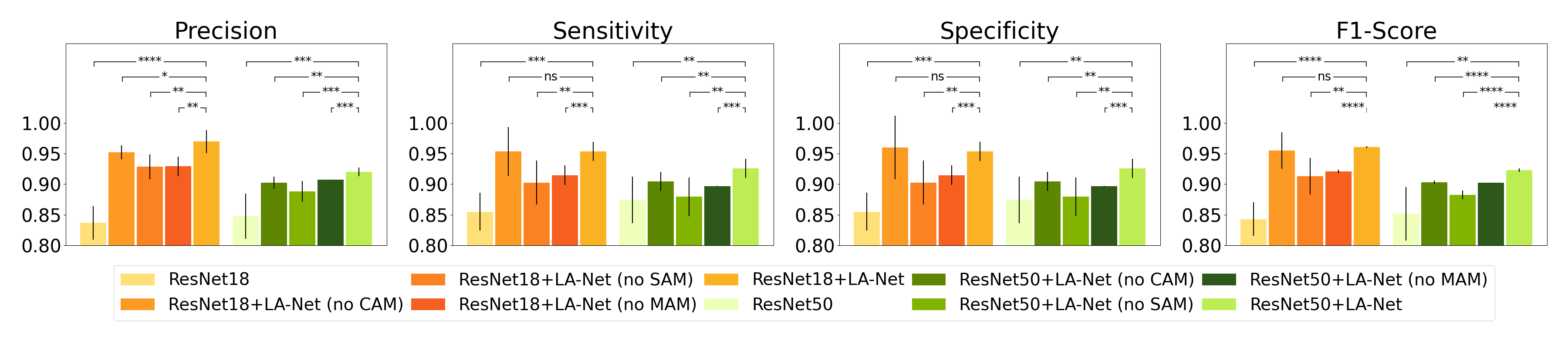}
        \caption{Image size: 512}
        \label{subfig:abl_mayo_512}
    \end{subfigure}
    \caption{Ablation study to investigate the effect of attention modules on classification performance of the proposed method on the F-MBUD dataset. }
    \label{fig:abl_mayo}
\end{figure}

The effect of each attention module on the classification performance based on the BUD dataset is shown in Figure~\ref{fig:abl_busi}.
The classification performance achieved by our method with all three attention modules was considered as the baseline.
When the input image size was $256\times256$ pixels, using all three attention modules achieved superior performance and
removing CAM, SAM, or MAM led to performance degradation compared to the baseline.
Particularly, removing SAM achieved the largest degradation, indicating the importance of SAM in improving classification performance by distilling lesion spatial knowledge.
When the input image size was up-scaled to $512\times 512$ pixels,
similar phenomena were observed by use of ResNet18 as FEX.
On the contrary, removing all three attention modules didn't show significant degradation when using ResNet50 as FEX.
It implies that the degree of improvement achieved by these attention modules is related to network factors, including FEX architecture and input image size.
Likewise, similar trends were also observed by use of F-MBUD dataset, as shown in Figure~\ref{fig:abl_mayo}, which confirms the significance of combining all three attention modules and highlights the impact of SAM.
These results validate the three attention modules make different contributions to improving model performance and the effect of SAM is highlighted.

\begin{figure}[!ht]
    \centering
    \begin{subfigure}[b]{0.5\textwidth}
        \includegraphics[width=\textwidth]{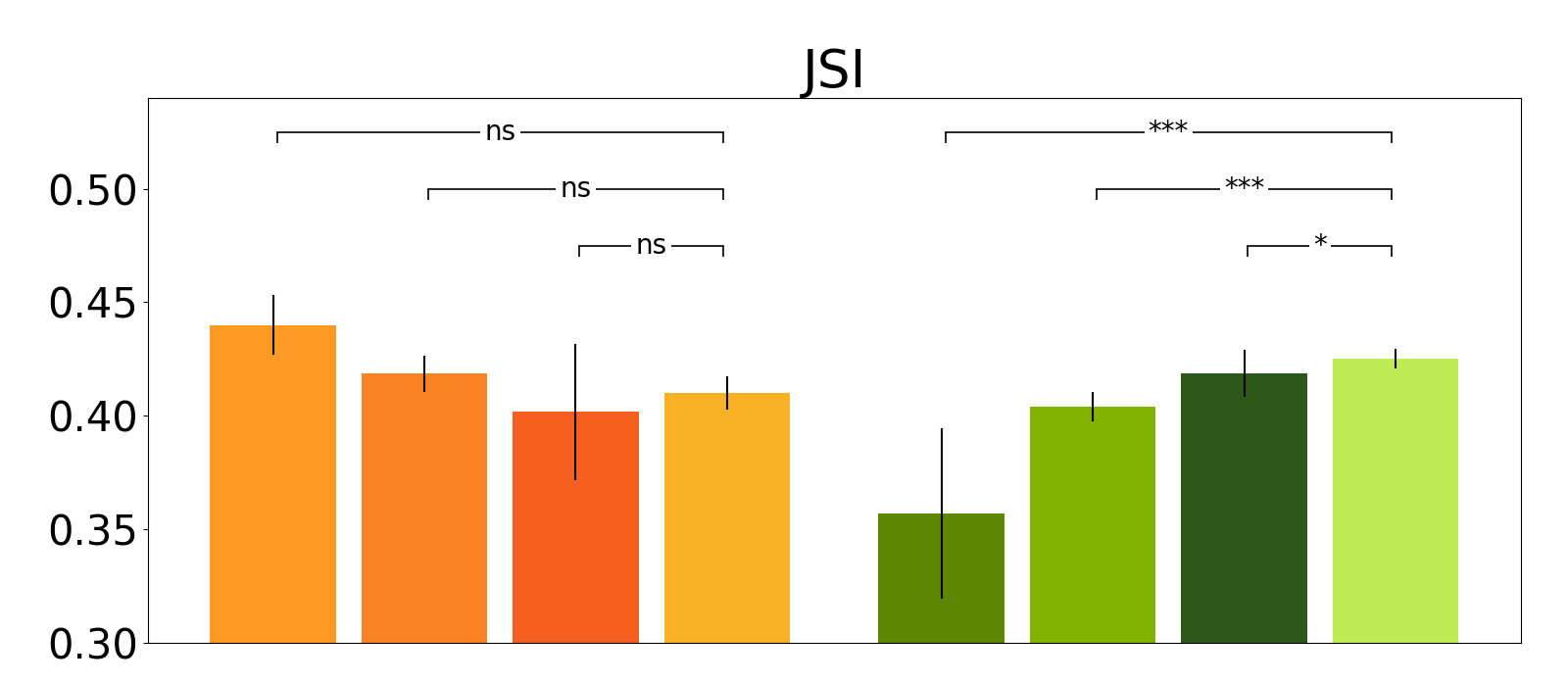}
        \caption{Image size: 256}
        \label{subfig:abl_busi_iou_256}
    \end{subfigure}
    \hfill
    \begin{subfigure}[b]{0.5\textwidth}
        \includegraphics[width=\textwidth]{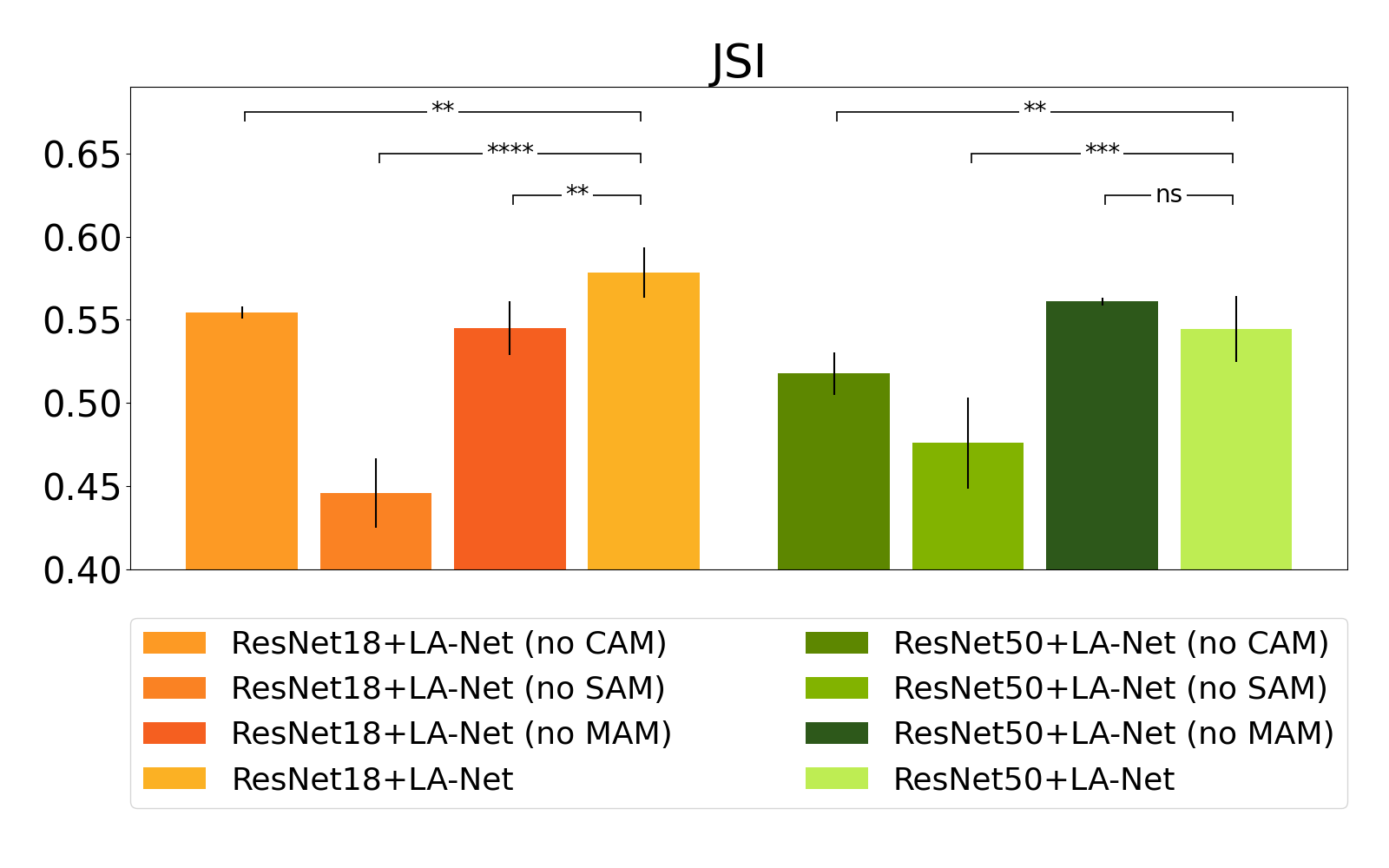}
        \caption{Image size: 512}
        \label{subfig:abl_busi_iou_512}
    \end{subfigure}
    \caption{Ablation study to investigate the effect of attention modules on localization performance of the proposed method on the BUD dataset. }
    \label{fig:abl_busi_iou}
\end{figure}

\begin{figure}[!ht]
    \centering
    \begin{subfigure}[b]{0.5\textwidth}
        \includegraphics[width=\textwidth]{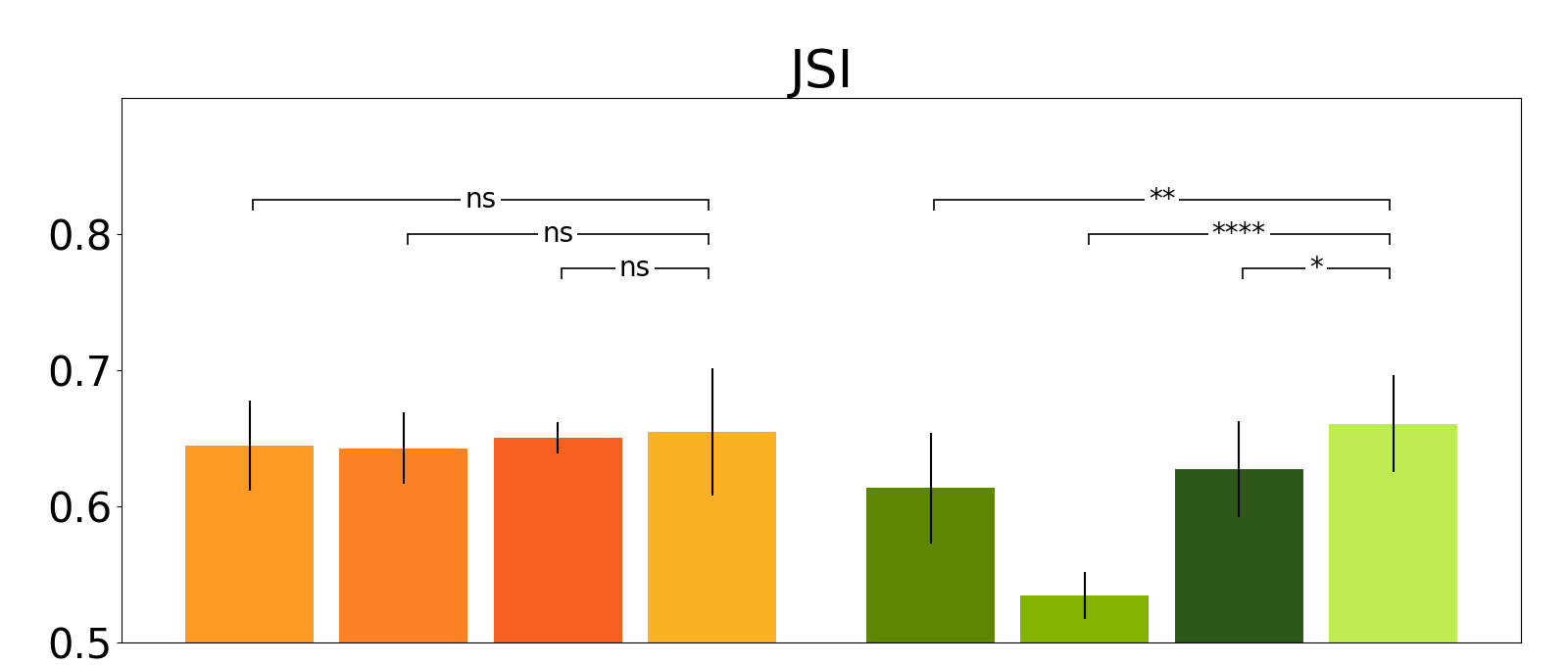}
        \caption{Image size: 256}
        \label{subfig:abl_mayo_iou_256}
    \end{subfigure}
    \hfill
    \begin{subfigure}[b]{0.5\textwidth}
        \includegraphics[width=\textwidth]{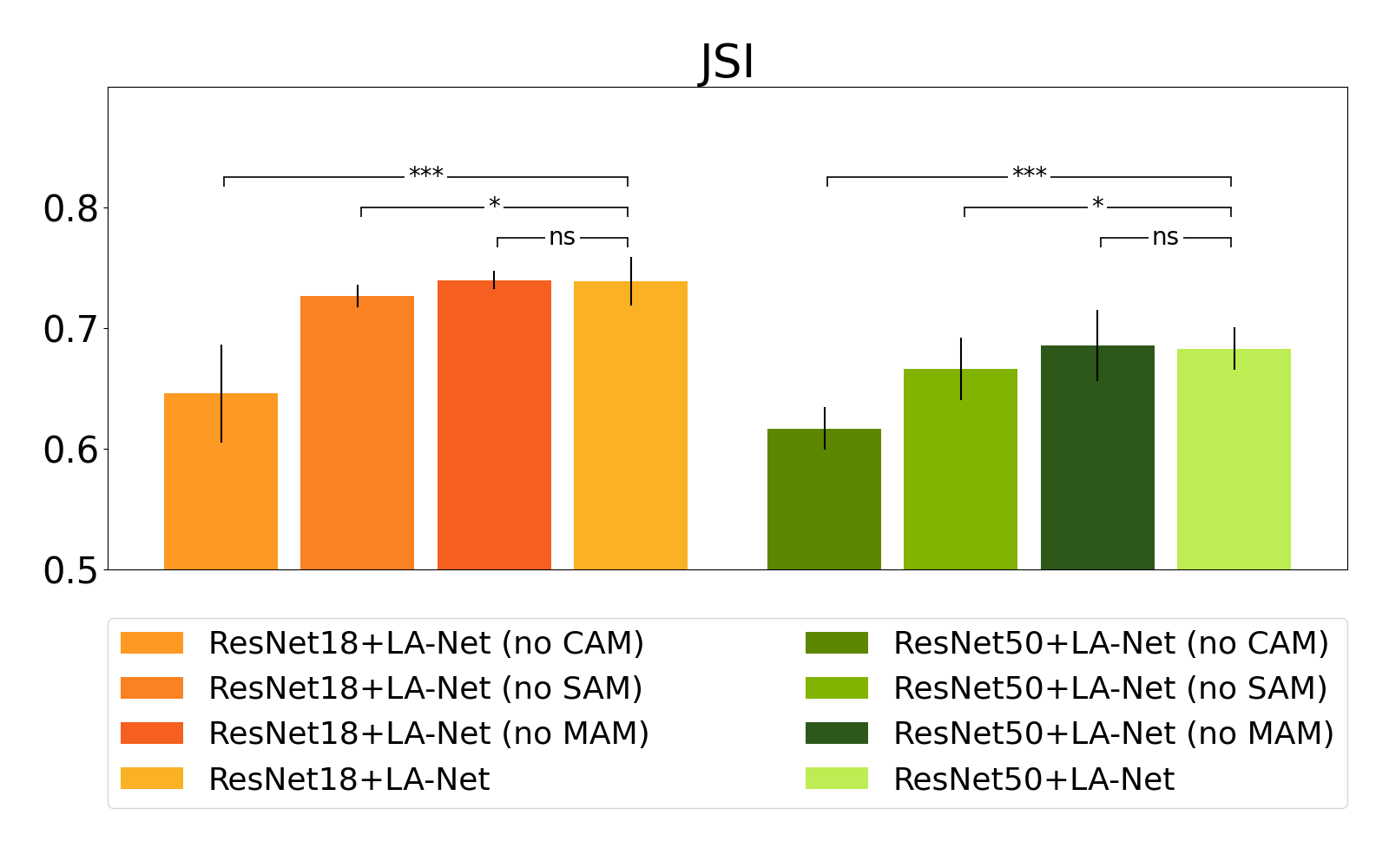}
        \caption{Image size: 512}
        \label{subfig:abl_mayo_iou_512}
    \end{subfigure}
    \caption{Ablation study to investigate the effect of attention modules on localization performance of the proposed method on the F-MBUD dataset. }
    \label{fig:abl_mayo_iou}
\end{figure}

Considering the localization performance, similar phenomena were observed based on both BUD dataset and F-MBUD dataset, as shown in Figure~\ref{fig:abl_busi_iou} and Figure~\ref{fig:abl_mayo_iou}, respectively.
Likewise, removing CAM or SAM significantly degraded the improvement when the image size was $512\times512$ pixels, while SAM still showed the dominant role.
However, no performance improvement was observed by using any attention module when ResNet18 was FEX and image size was $256\times256$ pixels.
These results confirm that the contribution of each attention module to the localization performance also varies according to network-related factors, while the importance of SAM is highlighted.

\subsection{Effect of incomplete datasets on the proposed method}
\label{subsec:part_data}
\begin{figure}[!ht]
    \centering
    \begin{subfigure}[b]{0.9\textwidth}
        \includegraphics[width=\textwidth]{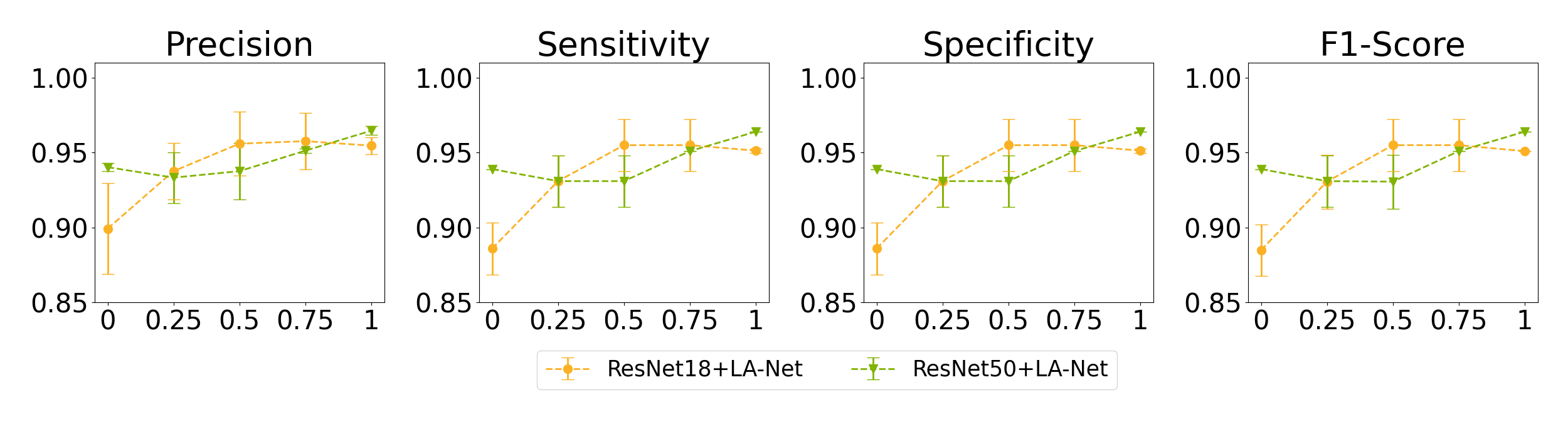}
        \caption{Image size: 256}
        \label{subfig:partial_busi_256}
    \end{subfigure}
    \hfill
    \begin{subfigure}[b]{0.9\textwidth}
        \includegraphics[width=\textwidth]{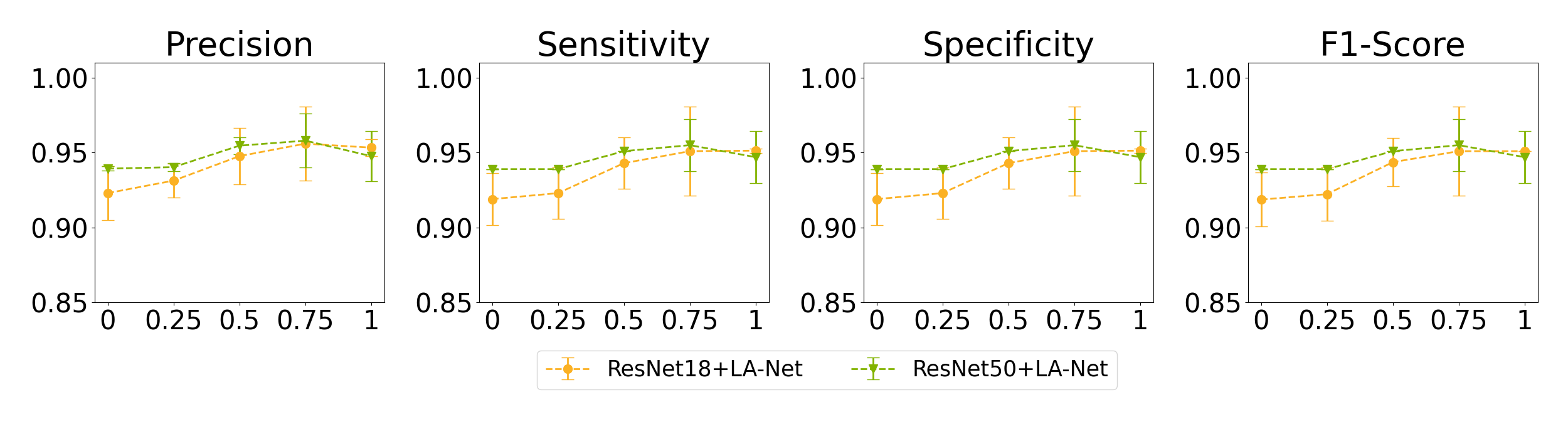}
        \caption{Image size: 512}
        \label{subfig:partial_busi_512}
    \end{subfigure}
    \caption{Classification performance of the proposed method on BUD-derived datasets with different ratios of lesion location labels.
    The percentage of GT location labels is shown on the x-axis, while zero means using vanilla ResNets for classification.
    Each solid point represents the mean of each metric under the given annotation ratio and the shadow shows its 95\% confidence interval.}
    \label{fig:net_effect_partial_busi}
\end{figure}

\begin{figure}[!ht]
    \centering
    \begin{subfigure}[b]{0.9\textwidth}
        \includegraphics[width=\textwidth]{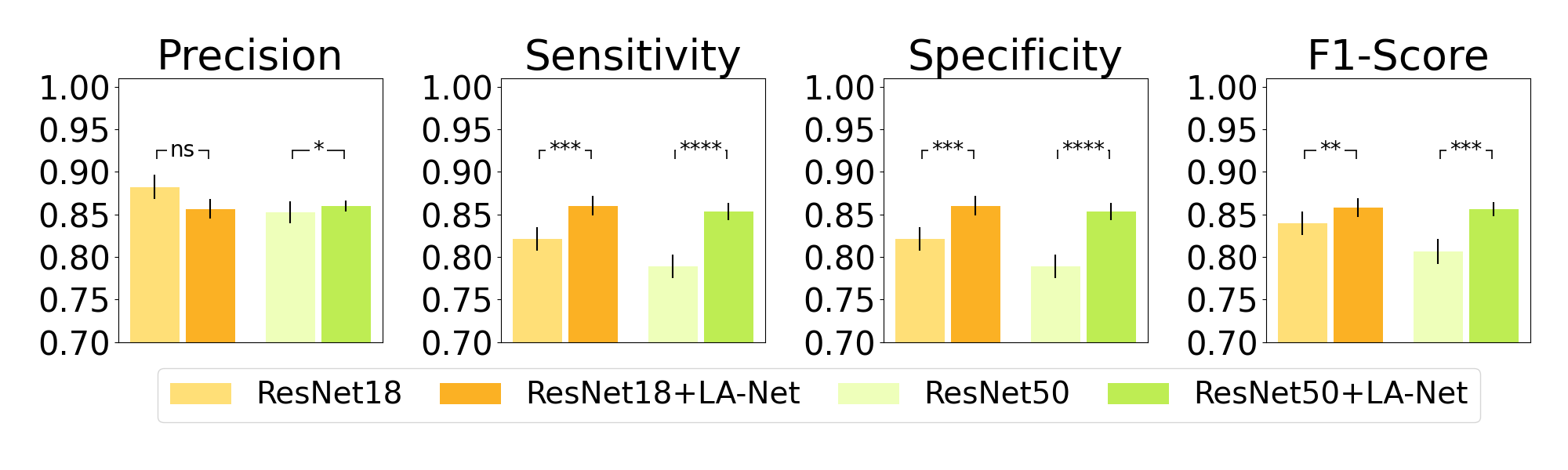}
        \caption{Image size: 256}
        \label{subfig:partial_mayo_256}
    \end{subfigure}
    \hfill
    \begin{subfigure}[b]{0.9\textwidth}
        \includegraphics[width=\textwidth]{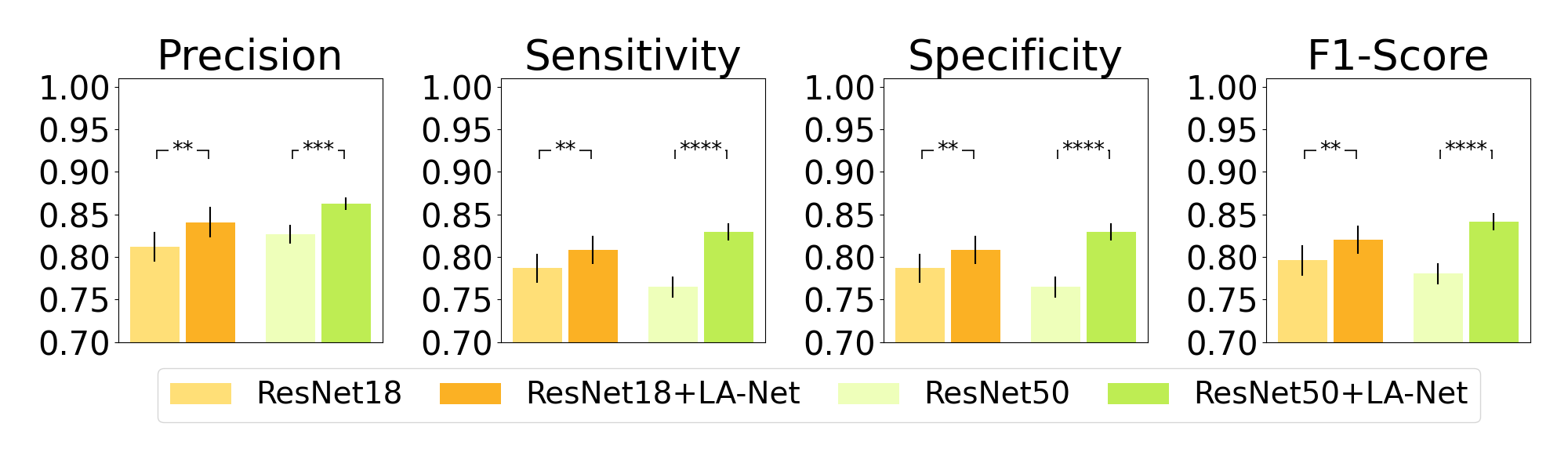}
        \caption{Image size: 512}
        \label{subfig:partial_mayo_512}
    \end{subfigure}
    \caption{Classification performance of the proposed method on the P-MBUD dataset. }
    \label{fig:net_effect_partial_mayo}
\end{figure}

The disentangled semi-supervised training strategy described in Section~\ref{subsec:train} enables our model to be trained on weakly-annotated datasets with incomplete and inaccurate location labels.
To verify this effect, incomplete datasets with partially-annotated location labels were employed, including P-MBUD dataset and BUD-derived incomplete datasets where the incomplete BUD dataset was prepared by keeping 25\%, 50\% or 75\% location labels and discarding the rest.
As shown in Figure~\ref{fig:net_effect_partial_busi}, compared to the vanilla ResNet, our method generally achieved significantly better classification performance on incomplete BUD datasets, showing the benefit of using localization annotation for the classification task.
Using more localization labels generally leads to better classification performance.
Interestingly, using 50\% localization labels can achieve comparable performance as that using the whole dataset, which implies incomplete annotation might not always lead to worse performance.
Similar results were observed based on P-MBUD dataset, as shown in Figure~\ref{fig:net_effect_partial_mayo}.
These results show the effectiveness of our proposed method in improving classification performance even on an incomplete training dataset.

\begin{figure}[!ht]
    \centering
    \begin{subfigure}[b]{0.49\textwidth}
        \centering
        \includegraphics[width=0.75\textwidth]{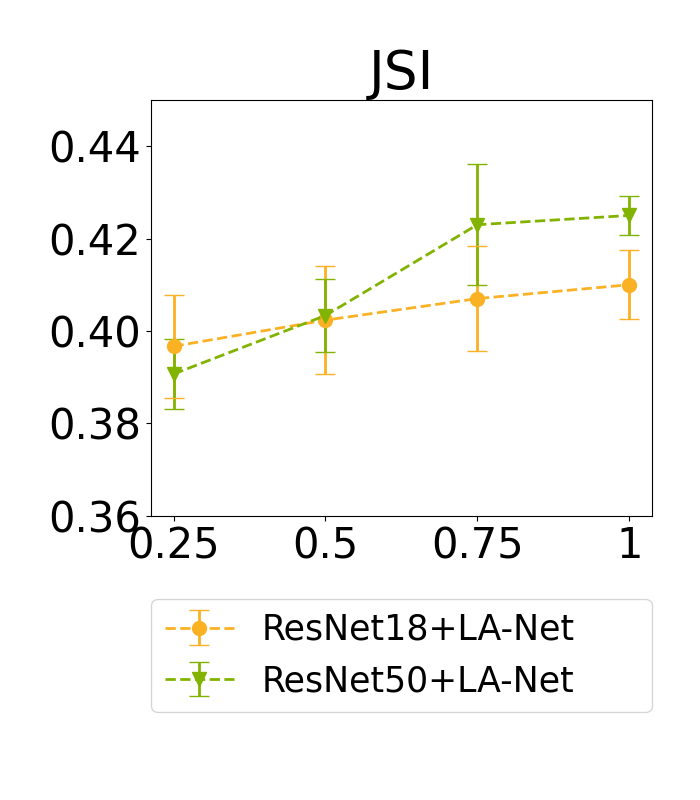}
        \vspace{-0.8cm}
        \caption{Image size: 256}
        \label{subfig:partial_busi_iou_256}
    \end{subfigure}
    \hfill
    \begin{subfigure}[b]{0.49\textwidth}
    \centering
        \includegraphics[width=0.75\textwidth]{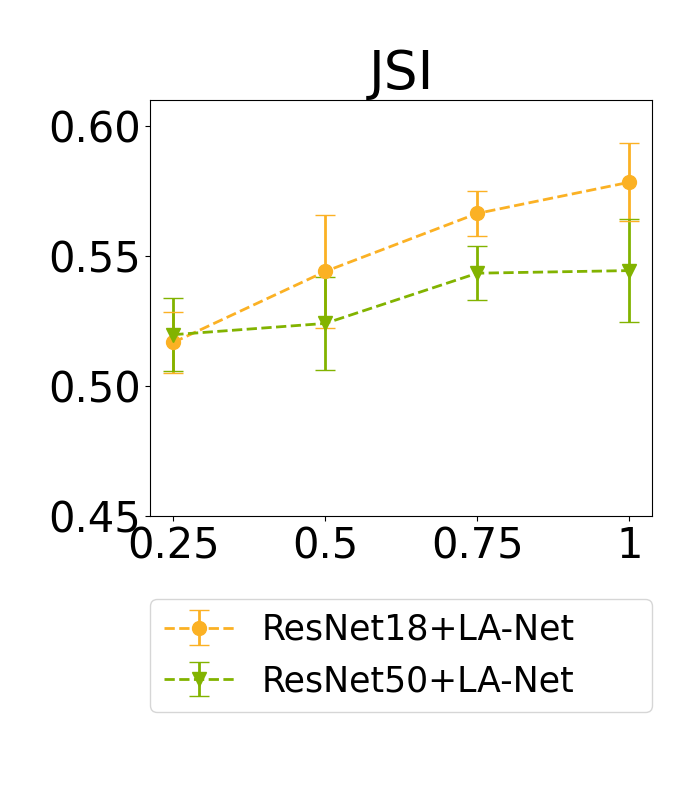}
        \vspace{-0.8cm}
        \caption{Image size: 512}
        \label{subfig:partial_busi_iou_512}
    \end{subfigure}
    \caption{Localization performance of the proposed method on the BUD-derived partially-segmented datasets (25\%, 50\% and 75\%).
    }
    \label{fig:net_effect_partial_busi_iou}
\end{figure}

The lesion localization performance of the proposed method on incomplete BUD datasets is shown in Figure~\ref{fig:net_effect_partial_busi_iou}. 
Likewise, using more localization labels leads to better localization performance, while 75\% localization labels can achieve similar performance as using the whole labels.
Noticeably, using ResNet50 as FEX performed better than that using ResNet18 when the image size was $256\times256$ pixels, but the trend was reversed when the image size was $512\times512$ pixels.
It shows the impact of network factors like FEX architecture and input image size on the semi-supervised localization performance.
The results trained on P-MBUD dataset are shown in Table~\ref{tab:iou_mayo_partial}. 
Our method achieved around 0.5 of JSI metric using the incomplete training datasets, representing a promising localization performance.
These results show the advantage of our method over other traditional MTL methods, which usually need datasets with fully-segmented annotations. 

\begin{table}[!ht]
    \centering
    \begin{tabular}{c | c | c }
        \hline
        Network & Image size & JSI \\
        \hline
        \multirow{2}{*}{ResNet18+LA-Net} & 256 & $0.459\pm 0.007$ \\
        & 512 & $0.569\pm 0.011$ \\
        \hline
        \multirow{2}{*}{ResNet50+LA-Net} & 256 & $0.531\pm 0.024$ \\
        &512 & $0.542\pm 0.021$ \\
        \hline 
    \end{tabular}
    \caption{Localization performance of the proposed method on P-MBUD dataset. The values are represented as mean value $\pm 95\%$ confidence interval.}
    \label{tab:iou_mayo_partial}
\end{table}

\subsection{Classification result interpretation by visualizing the discriminative region}
\begin{figure}[!ht]
    \centering
    \includegraphics[width=0.7\textwidth]{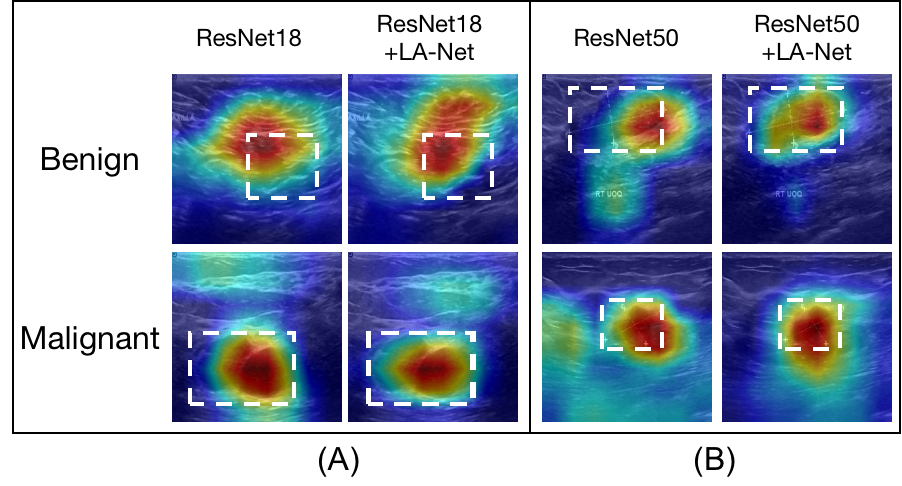}
    \includegraphics[width=0.7\textwidth]{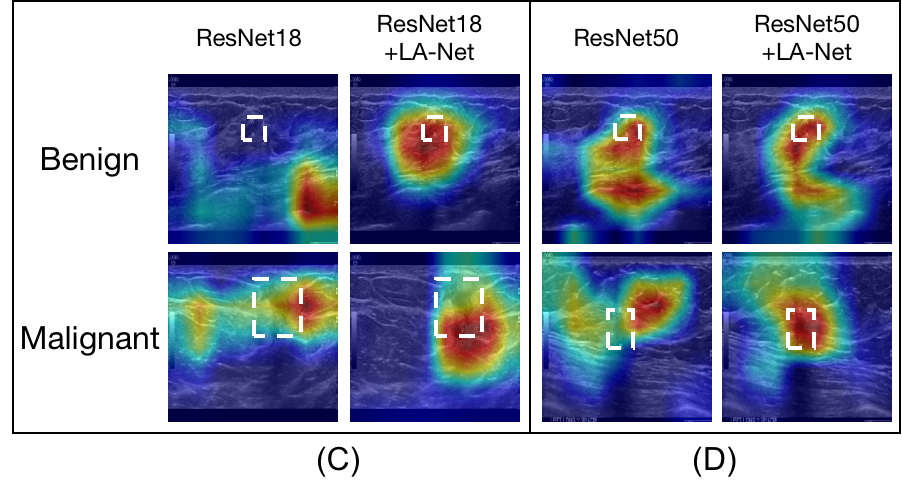} 
    \caption{Saliency map visualization for the last convolutional layer of FEX with/without LA-Net on BUD dataset (A and B) and F-MBUD dataset (C and D). The white dashed rectangles indicate the GT lesion locations. Red regions on the heatmaps correspond to highly confident regions of the predicted class labels.}
    \label{fig:saliency}
\end{figure}

To improve model interpretability, class activation mapping (CAM)~\cite{zhou2016learning} was proposed to describe how a DL-based model predicts the outcome by identifying discriminative regions on a given image and employed broadly to interpret classification networks in BUS image analysis field~\cite{Byra2022ExplainingAD,ding2022joint}.
Particularly, Grad-CAM, an effective generalization of CAM which is applicable for a variety of CNN models, can provide class-discriminative localization for visual explanations~\cite{selvaraju2017grad}.
It computes the gradient score for each class in terms of the feature map activation of the last convolutional layer in a CNN model, which renders high-level semantic and spatial information.
In this study, Grad-CAM was employed to highlight the potential ROIs based on the extracted feature of the last convolutional layer of FEX,
representing the learned spatial information for classification.
As shown in Figure~\ref{fig:saliency}, the attention regions recognized by our method have higher overlaps with the GT lesion locations.
It indicates that our model design enables FEX to focus on the lesion location more precisely during feature extraction, thus reducing the interference from the noisy background and improving the classification performance.

\section{Discussion}
\label{sec:disc}

In this study, a novel MTL framework is proposed for joint localization and classification of breast tumors on BUS images by use of a disentangled semi-supervised learning strategy and multiple attention modules.
Our proposed method shows three major benefits compared to traditional MTL methods.
First, a lesion-aware network is designed to alleviate the potential information sharing conflicts by use of multiple attention modules.
Second, a disentangled semi-supervised learning strategy enables the model to be trained on weakly-annotated datasets with incomplete lesion location labels, greatly reducing the burdens on data annotation.
Also, the model is modularized so that each network branch can be flexibly replaced with suitable architectures to satisfy various requirements.

The experiment results performed on the two BUS image dataset have shown the effectiveness of our method in improving model performance in both classification and localization tasks.
The performance of the proposed method is influenced by various network-related factors, including the architecture complexity of the feature extractor and the input image size.
According to the ablation study, three attention modules utilized in the framework show different contributions to performance improvement.
Particularly, the benefit of the spatial attention module is highlighted in both classification and localization tasks.
In addition, by use of the semi-supervised learning strategy, training the MTL model on incomplete datasets with partial location labels may not necessarily lead to degraded performance.

In the future, we would focus on further developing and modifying our network architecture to achieve better performance and to be generalized to various applications. 
More attention mechanisms to alleviate the potential information sharing conflict and fit more data modalities.
For example, when extending the method for a 3-dimension (3D) BUS scanning video dataset, temporal attention can be introduced to capture the temporal inter-frames correlation information~\cite{yan2019stat}.
More advanced semi-supervised learning techniques would be investigated to further reduce the demand for data annotation, such as generative-based methods~\cite{zhou2018brief,Yang2021ASO,zhai2022ass} and consistency regularization methods~\cite{Mittal2021SemiSupervisedSS}.
In addition, GAN-based adversarial learning strategies would be investigated to be integrated into our model to improve the lesion localization performance~\cite{singh2020breast,Fan2021DeepLM}.
Utilizing modern network architectures like EfficientNet~\cite{Tan2019EfficientNetRM} as the feature extractor can be investigated for better model performance and robustness.
Another interesting direction is to explore MTL frameworks using multi-modal ultrasound data, such as color Doppler and shear wave~\cite{chang2013comparison}, in order to improve both classification and localization performance with the rich input information. 
Apart from exploring methods for better model performance, we would also focus on investigating more advanced model interpretability methods, such as counterfactual explanation~\cite{Goyal2019CounterfactualVE} and concept whitening~\cite{Chen2020ConceptWF}. 
Explainable models can help provide transparent and safe predictions to aid clinicians in making better decisions in diagnosis.

\section{Conclusions}
\label{sec:conclusion}
A novel MTL framework is proposed for joint breast tumor classification and localization on BUS images.
Three attention modules are integrated into the framework to make the extracted feature more representative for better classification and localization performance.
A disentangled semi-supervised training strategy enables the training of the model on incomplete datasets.
This modularized framework holds the potential to flexibly choose suitable architecture for classification network and lesion-aware network to fit various clinical applications.
Experiments on two BUD datasets were employed to demonstrate the effectiveness of our method in terms of both classification and localization metrics.

\section{Acknowledgment}

This work was supported by NIH Award R01CA233873, 
Department of Defense (DoD) through the Breast Cancer Research Program (BCRP) under Award No. E01 W81XWH-21-1-0062,
and Cancer Center at Illinois Seed Grant. 
Opinions, interpretations, conclusions and recommendations are those of the author and are not necessarily endorsed by the Department of Defense.
Conflict of interest: none declared. 

\vspace{-0.2cm}
\bibliographystyle{ieeetr}
\bibliography{ref}

\end{document}